\documentclass[aps,twocolumn,showpacs,preprintnumbers,nofootinbib,prd,superscriptaddress,groupedaddress,10pt]{revtex4-1}

\usepackage{graphicx,amssymb,amsmath,amsthm,amsfonts,mathrsfs,dsfont,epsfig,epsf}
\usepackage[linktocpage]{hyperref}
\usepackage[usenames]{color}
\usepackage{epstopdf}
\usepackage{aas_macros}
\usepackage{bm}
\usepackage{dcolumn}
\usepackage[latin1]{inputenc}
\usepackage{latexsym}
\usepackage{rotating}
\usepackage{hyperref}
\usepackage{color}
\usepackage{longtable}
\usepackage{enumerate}
\usepackage{tensor,multirow}
\usepackage{url,float}
\def\be{\begin{equation}}
\def\ee{\end{equation}}
\newcommand{\beq}{\begin{eqnarray}}
\newcommand{\eeq}{\end{eqnarray}} 
\def\ba{\begin{align}}
\def\ea{\end{align}}

\begin{document}

\title{The Penrose process, superradiance and ergoregion instabilities}

\author{
Rodrigo Vicente$^{1}$,
Vitor Cardoso$^{1,2}$,
Jorge C. Lopes$^{1}$
}
\affiliation{${^1}$ CENTRA, Departamento de F\'{\i}sica, Instituto Superior T\'ecnico -- IST, Universidade de Lisboa -- UL,
Avenida Rovisco Pais 1, 1049 Lisboa, Portugal}
\affiliation{${^2}$ Perimeter Institute for Theoretical Physics, 31 Caroline Street North
Waterloo, Ontario N2L 2Y5, Canada}
\begin{abstract}
Superradiant scattering is a radiation enhancement process that takes place in many contexts, and which has recently found exciting applications in astro and particle physics.
In the framework of curved spacetime physics, it has been associated with the classical Penrose process for particles.
Superradiance is usually also associated with \textit{bosonic} fields around geometries with ergoregions and \textit{horizons}.
These notions are in clear tension however: the Penrose process occurs for \textit{horizonless} geometries, and particles are composed of \textit{fermions}.

Here, we resolve the tension in its different aspects, by showing that (i) superradiance occurs for self-interacting fermions on flat spacetime; (ii) superradiance occurs also for horizonless geometries, 
where it leads to an ergoregion instability. Ultracompact, horizonless geometries will usually respond with echoes of growing amplitude,
until rotational (or electrostatic) energy is extracted from the object;
(iii) the Fourier-domain analysis leads to absence of superradiance when horizons are not present. We elucidate why this analysis fails to give meaningful results;
(iv) finally, we show that superradiant, ergoregion instabilities have a particle analog of similar growth timescales and which can power the formation of a structure outside a compact, rotating star.
\end{abstract}
\maketitle

\tableofcontents

%
%
%
    %
%
\clearpage
\newpage

\begin{widetext}
\section*{Notation and conventions}
In this paper we use the metric signature $(+ \; - \; - \; -)$ . We also use units with $c= \hbar= G=k_e=1$ where $c$ is the speed of light in vacuum, $\hbar$ is the normalized Planck constant, $G$ is the gravitational constant and $k_e$ is the Coulomb constant.  For reference, the following is a list of
symbols that are used often throughout the text.
\begin{table}[h]
\begin{tabular}{ll}
  $g_{\alpha \beta}$ & Spacetime metric; Greek indices run from 0 to $3$. \\
  $e_a^{\,\,\,\mu}$  & Orthonormal spacetime tetrad. \\
  $M$                & Black hole mass. \\
	$Q$                & Black hole charge. \\
	$J$                & Black hole angular momentum.\\
	$a\equiv J/M$                & Angular momentum per unit of mass.\\
  $r_+$              & Radius of the BH event horizon in the chosen coordinates.\\
	$\Omega_+$         & Angular velocity ``of the horizon'' defined as $\Omega_+=a/(2Mr_+)$.\\
	$\Phi_+$         & Electric potential at the horizon.\\
$\mathds{1}$         & Identity matrix/operator.\\
$\nabla_\mu$         & Levi-Civita covariant derivative.\\
$Y_l^{m}$            & Spherical scalar harmonic of degree $l$ and order $m$ where $l$ and $m$ are integers satisfying $l\geq|m|$.\\
                     & These functions satisfy the orthonormality relations $\int d\Omega \, (Y_l^{m})^* Y_{l'}^{m'}= \delta_{l, l'}\, \delta_{m, m'}$.\\
$\chi_{j\mp1/2}^{k}$ & Spinor spherical harmonics  where $j$ and $k$ are half-integers satisfying $j\geq|k|$.\\
$\sigma^i$           & Pauli matrices,	$\sigma^1= \left( \begin{array}{cc}
	0 & 1 \\
	1 & 0
	\end{array} \right)$, $\sigma^2= \left( \begin{array}{cc}
	0 & -i \\
	i & 0
	\end{array} \right)$, $\sigma^3= \left( \begin{array}{cc}
	1 & 0 \\
	0 & -1
	\end{array} \right)$. \\
	$u_\pm$			& Normalized eigenvectors of Pauli's matrix $\sigma^3$ satisfying $\sigma^3 u_\pm=\pm u_\pm$ and $(u_\pm)^\dagger u_\pm=1$. \\
	$\gamma^a$       & Dirac $4\times4$ matrices $\gamma^0=\left( \begin{array}{cc}
	\mathds{1}_2 & 0 \\
	0 & -\mathds{1}_2
	\end{array} \right)$, $\gamma^i=\left( \begin{array}{cc}
	0 & \sigma^i \\
	-\sigma^i & 0
	\end{array} \right)$, $i=1,\,2,\,3 \,.$\\
	$G^\mu \equiv \gamma^a e_a^{\,\,\,\mu}$    & Curved spacetime Dirac matrices.\\
	$\phi$                & Klein-Gordon field.\\
	$A_{\mu}=(\Phi,\vec{A})$   & Electromagnetic potential.\\
	$\Psi$                     & Dirac spinor.\\
	$\bar{\Psi}=\Psi^\dagger G^0$ & Conjugate of Dirac spinor. In flat spacetime $G^0=\gamma^0$.\\
	$\Phi$                     & Electric potential, the time component of $A_{\mu}$.\\
	$A^*$                      & Complex conjugate of $A$.\\
	$A^\dagger$                & Hermitian conjugate of $A$.\\
  $\omega$                   & Fourier transform variable. The time dependence of any field is $\sim e^{-i\omega t}$.  \\
	$k, s$                        & Wavenumber. \\
  $\varphi_\omega \equiv \int e^{i\omega t}\phi(t,z) dt$& Fourier-transform of field $\phi$.\\
  $\omega_R,\,\omega_I$      & Real and imaginary part of the quasinormal modes (QNM).\\
	$q$                        & Charge of the fields.\\
	$\mu$                      & Mass of the fields.\\
 $I, R, T$                     & Amplitude of the incident, reflected and transmitted waves, respectively.\\
                               &These amplitudes are complex-valued functions of $\omega$. \\
\end{tabular}
\end{table}
\end{widetext}


%
\bigskip
\clearpage
\newpage

\section{Introduction}

Superradiance is a phenomenon where radiation is enhanced, and occurs in several contexts, from fluid dynamics~\cite{Cardoso:2016zvz}, quantum optics \cite{Dicke,GROSS1982301}, quantum mechanics \cite{MANOGUE1988261,Greiner1985} to electromagnetism~\cite{Bekenstein:1998nt,Yakov}. Other, perhaps more familiar, examples include Cherenkov 
radiation or the scattering of sound waves off supersonic interfaces~\cite{Brito:2015oca,Cardoso:2016zvz}. Superradiance can also be thought of as a friction mechanism that suppresses translational or rotational motion. As such, a coupling between radiation and the system providing the energy must exist in order for superradiance to be triggered. An interesting manifestation of this phenomenon occurs in the scattering of fields by these systems, 
and gives rise to larger amplitude scattered waves: there is a transfer of energy between the system and the radiation, the system looses energy that is transferred via the coupling mechanism to the field~\cite{Brito:2015oca}. 

\subsection{The spin of superradiance}
It is a fact of nature that all known particles fall into one of two big families: fermions (particles with half-integer spin) and bosons (particles with integer spin). Quarks and leptons are fermions, while the force carrier particles are bosons. The main difference between these two families is that fermions obey the Pauli exclusion principle which states that two identical fermions cannot be in the same state at the same point of spacetime. The strength of superradiant scattering of fields depends on which family the field belongs to~\cite{Brito:2015oca,Kim}. 
In the context of curved spacetime, the study of superradiance by black holes (BHs) started in 1971 with independent predictions by Zel'dovich and Misner that some waves could be amplified by rotating (Kerr) BHs~\cite{Yakov,Misner:1972kx}. A quantitative analysis became possible when linearized fields in these geometries were separated and decoupled~\cite{Teukolsky:1972my}. Using these tools, Teukolsky and Press proved that scalar, electromagnetic and gravitational waves scattering on a Kerr BH have superradiant modes~\cite{Teukolsky:1974yv,Brito:2015oca}. In fact, for massive bosons superradiance might trigger an instability that spins the BH down and produces characteristic, monochromatic gravitational waves and peculiar spin distributions~\cite{Arvanitaki:2016qwi,Brito:2014wla,Brito:2017wnc,Brito:2017zvb}. The first direct detection of gravitational waves from the \textit{GW150914} event by LIGO~\cite{Abbott:2016blz}, most likely originating from a pair of merging BHs (but see Refs.~\cite{Cardoso:2016oxy,Cardoso:2016rao}), contributes to making superradiance from BHs a potentially observable effect~\cite{Arvanitaki:2016qwi,Brito:2014wla,Brito:2017wnc,Brito:2017zvb,Rosa:2017ury}.

However, it is believed that the scattering of fermionic fields cannot be superradiant, due to the Pauli's exclusion principle. A rigorous and general mathematical proof of such statement is hard to come by, since
different fields obey different evolution equations. Nevertheless, the scattering of {\it free} Dirac spin-$1/2$ fields on a electrostatic potential barrier and on a Kerr-Newman (charged, rotating) BH cannot exhibit superradiance \cite{MANOGUE1988261,Greiner1985,Lee:1977gk,Unruh:1973bda,Chandrasekhar:1976ap,Lee:1977gk,Page:1976jj}. 

\subsection{Black holes, the Penrose process and superradiance} \label{subsec:penprocess}
We have alluded to superradiance from BHs, without commenting on the mechanism necessary to couple the fields to the geometry. Perhaps it is enlightening to consider Bekenstein's
argument for the existence of superradiance around BHs~\cite{Bekenstein:1973mi}. The argument is so simple and elegant that we outline it here. If the energy-momentum tensor of a (possibly charged) test field propagating on a Kerr-Newman background satisfies the null energy condition~\cite{Hawking:1973uf} at the event horizon, then the energy $\Delta M$, angular momentum $\Delta J$ and electric charge $\Delta Q$ absorbed by the BH satisfy~\cite{Natario:2016bay}:
\begin{equation} 
\label{areathm}
\Delta M \geq \Omega_+ \Delta J + \Phi_+ \Delta Q \,,
\end{equation}
where $\Omega_+$ is the angular velocity of the BH horizon and $\Phi_+$ is the electric potential at the horizon. It is easy to see that the ratio of the angular momentum or electric charge to the energy of a wave with frequency $\omega$, azimuthal number $m$ and electric charge $q$ are, respectively, $m/\omega$ and $q/\omega$~\cite{Bekenstein:1973mi}.
Then, the inequality \eqref{areathm} reads
\begin{equation}
\frac{\Delta M}{\omega} \left(\omega-m \Omega_+-q \Phi_+ \right) \geq 0 \,.
\end{equation}
Superradiant modes must extract energy from the BH and, so, $\Delta M<0$, which implies that $\omega$ must satisfy
\begin{equation}
0<\omega<m \Omega_++q \Phi_+ \,.
\end{equation}
These are, precisely, the modes which extract energy from the BH. Since the energy-momentum tensor of a Dirac field does not satisfy the null energy condition at the event horizon~\cite{Toth:2015cda}, such fields are not contemplated by this proof; in fact, as we said above, they do not exhibit superradiant amplification when scattering on this background.

Black holes are, by definition, solutions of the field equations that contain closed regions with their interior causally disconnected from their exterior; the surface separating these two causally separated regions is called the event horizon. Nothing (not even light) within the event horizon can influence outside observers~\cite{gravitation}. In other words, the horizon can be thought of as a ``one-way membrane.'' This uni-directionality provides a dissipative character to BH geometries \cite{Brito:2015oca,Richartz:2009mi}. One might thus be tempted to conclude that superradiance in BHs occur because horizons gives us the necessary coupling to the field:
most or all other examples of superradiance, like scattering from conducting materials, require some kind of friction or viscosity. Since BHs appear in several vacuum solutions of the Einstein field equations, the event horizon provides the vacuum itself with a dissipative mechanism. This is an interesting feature, because the fields are allowed to extract energy from the vacuum by superradiant scattering.

However, there is another ingredient in rotating BH backgrounds that could provide the necessary coupling, the ergosphere. In the ergoregion, matter is forced to co-rotate (as seen from infinity) with the BH, thus providing a very effective coupling to the spacetime. These two aspects of BHs, horizons and ergoregions, come hand in hand in General Relativity, and it seems hopeless to understand them separately.
However, a study of the scattering of monochromatic waves off potential barriers indicates that it is the boundary conditions on the fields that dictate superradiance~\cite{Richartz:2009mi}.
The results suggest that the absence of an horizon - for example, replacing the horizon by a hard surface - would lead to no superradiance at all~\cite{Richartz:2009mi}.

In a closely related development, in 1971, Roger Penrose theorized a process whereby rotational energy could be extracted from BHs
through the use of point particles~\cite{Penrose:1971uk,Brito:2015oca,gravitation}. 
The Penrose process makes use of the fact that particles (classical states of matter) in the ergoregion of Kerr BHs may have negative energy with respect to observers at infinity~\cite{Penrose:1971uk,Brito:2015oca,gravitation}. By letting a particle fall into the ergoregion and decaying there in two other particles, it is possible to carefully select the process such as to have, in the final state, one particle escaping to infinity with an energy larger than the one of the initial state. It can be shown that, for Reissner-Nordstr\"{o}m (charged, static) BHs, there exists a generalized ergoregion and a similar energy extraction process is possible~\cite{Denardo:1973pyo,1985JApA....6...85B}. 
In other words, classical energy extraction occurs for particles, and hinges solely on the existence of ergoregions (or, more generally, regions of negative energy).

\subsection{The tension}
Clearly, there is some tension between the elements above:

{\bf 1. Are horizons necessary?} The Penrose process does not require horizons, but superradiant scattering of monochromatic waves does, according to a stationary Fourier-domain analysis~\cite{Richartz:2009mi}. However, from a purely physical point of view,
superradiance should be independent of the exact boundary conditions at the horizon, since it takes an infinite amount of time for a wave to reach it. In fact, if the horizon at $r=r_+$
(in some coordinates) is replaced by a hard surface at $r=r_+(1+\delta)$, with $\delta\ll1$, then, clearly, the scattering should proceed as in the BH case. Differences should only be perceived after the wave has had time to probe the surface (which for Kerr geometries means for times $(\Delta t/M) \gtrsim |\text{Log}\delta|$, typically~\cite{Cardoso:2016oxy,Cardoso:2016rao,Cardoso:2017cqb,Cardoso:2017njb,Mark:2017dnq}).

{\bf 2. Are fermions superradiant-less?} As we said, the Penrose process is generally believed to be the particle analogue of superradiant scattering. However, while the two processes are classical, superradiant amplification seems to carry some quantum features of the field being scattered. In particular, even though the fields are not quantised, superradiant scattering already seems to carry some information on pair production~\cite{Brito:2015oca,MANOGUE1988261,Hansen:1980nc}. 

The Penrose process describes ordinary matter in curved spacetime. Thus, if one believes it to be the particle analogue of superradiance, then there is something to explain: ordinary (baryonic) matter is made of fermions at the very fundamental level, but fermions do not exhibit superradiant amplification, as we described above. If the two pictures are to be consistent, then either
quantum effects would quench the Penrose process or non-linear interactions between the fermions, not taken into account in any of the previous studies, would restore their capability to exhibit superradiance. It seems highly unlikely, and dangerous for known physics, that quantum effects would show up so dramatically in macroscopic systems. Thus, we are left with the second option.
 In other words, we expect the existence of bosonic fermion condensates with the capability to exhibit superradiant amplification. In fact, the existence of fermion systems with a bosonic behaviour is not very strange and happens in nature. For instance, Cooper pairs in BCS theory of superconductivity and mesons in particle physics are examples of these bosonic fermion condensates with built-in nonlinear interactions~\cite{Nambu}.

The existence of these kind of condensates could have interesting applications in astrophysics. It is known that we can confine superradiantly amplified fields through various mechanisms, like massive fields and anti-de Sitter boundaries~\cite{Brito:2015oca,Detweiler:1980uk,Cardoso:2004hs}. This confinement can originate strong instabilities called BH bombs~\cite{Press:1972zz,Cardoso:2004nk,VILENKIN1978301,Chesler:2018txn}, which have applications in searches for dark matter and physics beyond the Standard Model~\cite{Arvanitaki:2010sy,Brito:2015oca,Pani:2012vp,Brito:2013wya,Hod:2012px,Herdeiro:2014goa,Herdeiro:2015waa}.

\subsection{Plan}

First, we review the scattering of scalar and Dirac fields on electrostatic potential barriers (the Klein setup) and on Reissner-Nordstr\"{o}m (RN) BHs. The results obtained for these cases are well known~\cite{Greiner1985,MANOGUE1988261,Brito:2015oca,Lee:1977gk}. Nevertheless, as far as we know, the only proof of the absence of superradiance for Dirac fields on RN background is obtained as a limit of the more general Kerr-Newman background~\cite{Lee:1977gk}, which uses the formalism of Newman-Penrose~\cite{Newman:1961qr} to separate the Dirac equation. Here, we use the spherical symmetry of RN geometry to separate the Dirac equation in an easier way and we proceed to prove the absence of superradiance for Dirac fields on a RN background. Furthermore, we provide a non-linear Dirac field theory, which is inspired by the Nambu-Jona-Lasinio model~\cite{Nambu} and can exhibit superradiant amplification on the Klein setup. We are not concerned about the generality or physical validity of this theory. Instead, we want to provide a simple theory describing fermions which allows superradiant scattering solutions. In other words, ours is a proof of principle
that real-world particles, made of interacting fermions, could have superradiance {\it and} be subjected to the Penrose process as well. However, we show that the same theory does not admit superradiant solutions on RN geometry. Finding a theory that admits such solutions on curved spacetime does not seem to be a trivial problem. However, from a physical point of view, we expect such a theory to exist.

We proceed by doing an analysis of the scattering of wavepackets on the Klein setup and RN background, obtaining amplification factors which converge to the ones of monochromatic waves (in the quasi-monochromatic limit). We employ two different methods: a Fourier-domain calculation and a fully numerical time-domain analysis, giving consistent results. Then, we go on to prove that horizons (and, more generally, boundary conditions) have nothing to do with superradiance. In particular, we repeat the fully numerical time-domain analysis, this time using Dirichlet boundary conditions (simulating an hard surface which covers the horizon). The results prove that superradiance occurs also in this case, and the amplification factor is the same. But this time the scattering is accompanied by an ergoregion instability, characterized by the emission of echoes of growing amplitude. Moreover, we use this feature to elucidate why the Fourier-domain analysis fails to give meaningful results in the case of horizonless geometries.

Finally, we treat a simplified version of a Penrose process cascade in the ergoregion of a rotating, ultracompact star. The results show that there is a particle analog of the superradiant ergoregion instability, which is characterized by the same time scale.

\section{The framework}
We focus on charged test fields, on a background electromagnetic field and geometry. The test fields are taken to be so weak that backreaction on the electromagnetic potential or on the geometry is negligible. This test field approximation is correct at first order in the fields, because their effect on the geometry and on the electromagnetic field is only of second order~\cite{Brito:2015oca}. 
This approximation can in principle capture most, if not all, astrophysical scenarios.

The field theories that we work with throughout this work are described by an action of the form:
\begin{equation}
S=S_G+S_{E M}+S_M \,,
\end{equation}
with
\begin{align}
S_G&=\int d^4x \sqrt{-g}\,\, \frac{R}{16 \pi} \,, \\
S_{EM}&=-\int d^4x \sqrt{-g} \,\,\frac{1}{4} F_{\mu \nu} F^{\mu \nu} \,,
\end{align}
where $g$ is the determinant of the metric $g_{\mu \nu}$, $R$ is the scalar curvature and $F_{\mu \nu}$ is the electromagnetic field tensor
\begin{equation}
F_{\mu \nu}=\partial_\mu A_\nu- \partial_\nu A_\mu \,.
\end{equation}
The action $S_M$ describes the matter content. In this work, we will consider three kind of matter fields: scalar fields, Dirac fields and non-linear Dirac fields.

We focus on theories $S_M$ which are U(1) invariant. A theory of this kind is such that if it describes the field $\xi$, then its equations are invariant under the transformation $\xi \longrightarrow e^{i \alpha} \xi$
with $\alpha$ a real constant. Noether's theorem guarantees that there is a conserved current associated with this symmetry~\cite{Banados:2016zim}.  We call this current the particle-number current and we use its flux to study the phenomenon of superradiant scattering~\cite{Kim}. We say that there is superradiant amplification if the absolute value of the flux of the reflected particle-number current is larger than that of the incident one. 

\section{Klein setup}
In 1929, using the Dirac equation, Klein showed that an electron beam propagating in a region with a sufficiently large potential barrier can be transmitted without the exponential damping expected from non-relativistic quantum mechanics~\cite{Klein}. This phenomenon was dubbed Klein paradox by Sauter~\cite{Sauter:1931zz} and it can be explained by pair production at the potential barrier using quantum field theory~\cite{MANOGUE1988261,Calogeracos:1999yp,Brito:2015oca}. 
It is possible to show that for sufficiently large potential barriers there is a production of scalar and Dirac pairs, explaining the existence of transmitted modes instead of the exponential damping~\cite{MANOGUE1988261,Hansen:1980nc}. Also, it is known that superradiance occurs due to pair production; the absence of superradiant scattering for Dirac field can then be attributed to Pauli's exclusion principle~\cite{Brito:2015oca,MANOGUE1988261,Hansen:1980nc}. The behavior is fundamentally the same when the fields are scattering on a (possibly charged) Kerr background~\cite{Damour:1975pr,Arderucio:2014oua}.

Consider a two dimensional problem, depending on time and on the $z-$direction, on a four dimensional flat spacetime described by the line element $ds^2=dt^2-dx^2-dy^2-dz^2$ and by an electromagnetic potential
\begin{equation}
A_\mu=(\Phi(z),\,0,\,0,\,0) \,,
\end{equation}
with the asymptotic behaviour
\begin{equation} 
\label{asympot}
\Phi(z)= \begin{cases}
0, & \text{for} \,\, z\to -\infty \\
\Phi_0>0, & \text{for} \,\, z\to +\infty 
\end{cases} \,.
\end{equation}
We wish to see what happens to incident quasi-monochromatic charged waves coming from $z\to-\infty$ and scattering off this potential barrier. To interpret the waves as incident, reflected and transmitted, we use their group velocity. 

\subsection{Scattering of scalar fields}
Let us start by considering the scalar theory minimally coupled to an electromagnetic field,
\begin{equation} \label{scalaraction}
S_{\rm scalar}=\int dx^4\left[D_\mu\phi(D^\mu\phi)^*-\mu^2|\phi|^2\right] \,,
\end{equation}
with $\mu$ the mass of the field and with $$D_\mu =\partial_\mu +i q A_\mu \quad,$$ where $q>0$ is the electric charge of the field. 
From this action, one gets the equation of motion
\begin{equation} \label{scalarfe0}
(D^\mu D_\mu+\mu^2)\phi=0 \,.
\end{equation}
Using the ansatz $\phi_\omega(t,z)=e^{-i\omega t}\varphi_\omega(z)$, we find 
\begin{equation} \label{scalartife}
[\partial^2_z+(\omega-q \Phi)^2-\mu^2]\varphi_\omega=0 \,.
\end{equation}

This equation admits the asymptotic solution
\begin{equation} 
\label{scalariI}
\varphi_\omega(z\to-\infty)= I\, e^{i k z}+R\, e^{-i k z}\,,
\end{equation}
with
\be \label{drldks}
k= \epsilon \sqrt{\omega^2-\mu^2} \,,\qquad \epsilon=\text{sign}(\omega+\mu)\,.
\ee
We focus exclusively on propagating modes, thus $\omega^2>\mu^2$. The sign $\epsilon$ was chosen in such a way that makes the wave $e^{-i(\omega t- k z)}$ have group velocity
\begin{equation} \nonumber
	v_g=\left(\frac{d k}{d \omega}\right)^{-1}=\epsilon \frac{\sqrt{\omega^2-\mu^2}}{\omega}>0\,.
\end{equation}
Then, the $I$ and $R$ quantities are the complex-valued amplitudes of the incident and reflected waves, respectively.
The other asymptotic solution is
\begin{equation}
\label{scalariII}
\varphi_\omega(z\to +\infty)= T\,e^{i s z} \,,
\end{equation}
with
\be \label{drs}
s=\widetilde{\epsilon}\sqrt{(\omega-q\Phi_0)^2-\mu^2} \,,\qquad \widetilde{\epsilon}=\text{sign}(\omega-q\Phi_0+\mu)\,. 
\ee
Here, the sign $\widetilde{\epsilon}$ was chosen so as to make the wave $e^{-i(\omega t-s z)}$ have group velocity
\begin{equation}\nonumber
	v_g=\left(\frac{d s}{d \omega}\right)^{-1}=\widetilde{\epsilon} \frac{\sqrt{(\omega-q\Phi_0)^2-\mu^2}}{\omega-q\Phi_0}>0\,,
\end{equation}
for propagating transmitted modes (\textit{i.e.}, for $s\in \mathbb{R}$). Furthermore, in the case of non-propagating penetrating modes (\textit{i.e.}, $s\notin \mathbb{R}$), the chosen sign $\widetilde{\epsilon}$ implies
\begin{equation}\nonumber
	\text{Im}(s)=\widetilde{\epsilon}\sqrt{\mu^2-(\omega-q\Phi_0)^2}>0\,,
\end{equation}
making the stationary wave $e^{-i(\omega t-s z)}=e^{-i\omega t}e^{-s z}$ attenuated along $z$, and well-behaved at $z\to +\infty$.
Then, the quantity $T$ is the complex-valued amplitude of the transmitted (penetrating) solution. Both $(R/I)$ and $(T/I)$ are functions of $\omega$, with the functional form depending on the shape of the potential barrier.

Using the $z$-component of the particle-number current associated with the scalar field $\phi$,
\begin{equation} \label{pncurrentscalar}
j_{z}=-\frac{i}{2}(\phi^*\partial_{z}\phi-\phi\,\partial_{z}\phi^*) \,,
\end{equation}
we can show
\begin{equation} \label{scalaramp}
\left|\frac{j_z^r}{j_z^i}\right|=1- \frac{\text{Re}(s)}{k} \left|\frac{T}{I} \right|^2 \,,
\end{equation}
where $j_z^i$ and $j_z^r$ are the incident and reflected $z$-currents, respectively.

By definition, superradiance occurs when $|j_z^r|>|j_z^i|$, and is possible only if $q \Phi_0> 2 \mu$. The superradiant modes are
\begin{equation}
\mu< \omega < q \Phi_0-\mu \,.
\end{equation}
These modes are in agreement with the ones obtained in Ref.~\cite{MANOGUE1988261}.

\subsection{Scattering of Dirac fields}
Focus now on the theory of fermions,
\begin{equation} \label{Diracaction}
S_{\rm Dirac}=\int dx^4 (i \bar{\Psi} \gamma^{\mu} D_{\mu}\Psi-\mu\bar{\Psi}\Psi) \,,
\end{equation} 
with $\bar{\Psi}$ the Dirac conjugate of the spinor $\Psi$.
From this action, it follows the equation of motion
\begin{equation} \label{diracfe0}
i \gamma^{\mu}D_{\mu}\Psi-\mu\Psi=0 \,.
\end{equation} 
With the ansatz $\Psi_\omega(t,z)=e^{-i \omega t}\chi_\omega(z)$, where $\chi_\omega(z)$ is a 4-spinor, we find
\begin{equation} \label{diractife}
[i\gamma^3 \partial_z+(\omega-q \Phi)\gamma^0-\mu \mathds{1}
]\chi_\omega=0 \,.
\end{equation}
One of the asymptotic solutions of this equation is
\beq
&\chi_\omega&(z\to-\infty)\nonumber\\
&=&\left[I_+ \left(\begin{array}{c}
u_+ \\ \frac{k}{\omega+\mu}u_+
\end{array}\right) + I_- \left(\begin{array}{c}
u_- \\ -\frac{k}{\omega+\mu}u_-
\end{array}\right)\right] e^{i k z} \nonumber \\
&+& \left[R_+ \left(\begin{array}{c}
u_+ \\ -\frac{k}{\omega+\mu}u_+
\end{array}\right)+ R_- \left(\begin{array}{c}
u_- \\ \frac{k}{\omega+\mu}u_-
\end{array}\right)\right] e^{-i k z} \,,
\eeq
with $k$ satisfying Eq.~\eqref{drldks}. The $u_+$ and $u_-$ denote the normalized eigenvectors of Pauli's matrix $\sigma^3$ (\textit{i.e.}, $\sigma^3 u_\pm=\pm u_\pm$).
As in the scalar field case, we are considering modes with $\omega^2>\mu^2$. The $I_+$ and $I_-$ are the complex-valued amplitudes of the incident wave, and $R_+$ and $R_-$ the ones of the reflected wave.

The remaining asymptotic solution is
\begin{align}
\begin{split}
&\chi_\omega(z\to +\infty)\\&=\left[T_+ \left(\begin{array}{c}
u_+ \\ \frac{s}{\omega-q\Phi_0+\mu}u_+
\end{array}\right)+ T_- \left(\begin{array}{c}
u_- \\ -\frac{s}{\omega-q \Phi_0+\mu}u_-
\end{array}\right)\right] e^{-i s z} \,,
\end{split}
\end{align}
with $s$ satisfying Eq.~\eqref{drs}.
The $T_+$ and $T_-$ are the complex-valued amplitudes of the transmitted solution. Both $(R_+/I_+)$, $(R_-/I_-)$, $(T_+/I_+)$ and $(T_-/I_-)$ are functions of $\omega$.
Using the $z$-component of the particle-number current associated with the Dirac field $\Psi$,
\begin{equation} 
j^{z}=\frac{1}{2}\bar{\Psi}\gamma^{3} \Psi \,,
\end{equation}
one can show 
\be
\left|\frac{(j^r)^z}{(j^i)^z}\right|=1 - \frac{\omega+\mu}{k} \frac{\text{Re}(s)}{\omega-q \Phi_0+\mu} \frac{|T_+|^2 +|T_-|^2}{|I_+|^2 +|I_-|^2} \,, 
\ee
where $(j^i)^z$ and $(j^r)^z$ are the incident and reflected $z$-currents, respectively.
Thus, $(j^r)^z \leq (j^i)^z$: there are no superradiant modes of the Dirac field. This is in agreement with the results of Refs.~\cite{MANOGUE1988261,Greiner1985}.

\subsection{Scattering of non-linear Dirac fields}
For the reasons explained in the Introduction, we want to consider the scattering of a fermion condensate. We use the usual Dirac free field action with an additional interaction term involving $(\bar{\Psi}\Psi)^2$. This theory is sometimes referred to as the Soler model \cite{SOLER}. The extra term is such that the U(1) symmetry of $\Psi$ is preserved, giving rise to a Noether's conserved current.  The $z$-component of this current can be shown to be equal to the one of the free Dirac field. So, let us consider the non-linear Dirac field theory
\begin{equation}
S=\int dx^4 \left(i \, \bar{\Psi} \gamma^\mu D_\mu \Psi-\mu \bar{\Psi} \Psi +\frac{\lambda}{2}(\bar{\Psi}\Psi)^2 \right) \,,
\end{equation}
with
\begin{equation} \label{coupling}
\lambda(z)=\widetilde{\lambda}\, q^2 A_\mu A^\mu= \begin{cases}
0\,, &   z\to -\infty \\
\lambda_0=\widetilde{\lambda}\, q^2 \Phi_0^2 >0\,, &  z\to +\infty
\end{cases} \,,
\end{equation}
where $\widetilde{\lambda}>0$ is a real constant. Such coupling is motivated by the need to have linear equations of motion at $z \to -\infty$, 
in order to have a clear and well defined notion of ``incident'' and ``reflected'' wave.

This action yields the field equation
\begin{equation} \label{diracfe}
i \gamma^\mu D_\mu \Psi-\mu \Psi + \lambda (\bar{\Psi}\Psi) \Psi=0 \,,
\end{equation}
Using $\Psi_\omega(t,z)=N e^{-i\omega t}\chi_\omega(z)$, we obtain the time-independent field equation
\begin{equation} 
\label{diracnltife}
\left[i \gamma^3 \partial_z + (\omega - q \Phi)\gamma^0-\bigg(\mu-\lambda \bar{\chi}_\omega \chi_\omega\bigg) \mathds{1}\right]\chi_\omega=0 \,.
\end{equation}
Since the non-linear coupling vanishes at $z\to - \infty$, one asymptotic solution in this region is
\beq
&\chi_\omega&(z \to -\infty)\nonumber\\
&=&I_+ \left(\begin{array}{c}
u_+ \\ \frac{k}{\omega+\mu}u_+
\end{array}\right) e^{i k z}+ R_+ \left(\begin{array}{c}
u_+ \\ -\frac{k}{\omega+\mu}u_+
\end{array}\right) e^{-i k z} ,
\eeq
with $k$ satisfying Eq.~\eqref{drldks}.
Moreover, using the last section (\textit{i.e.} $\lambda_0=0$) as an inspiration, let us look for solutions with asymptotic behavior
\begin{equation} \label{transnl}
\chi_\omega(z\to +\infty)=T_+ \left(\begin{array}{c}
u_+ \\ \eta_\omega\, u_+
\end{array}\right) e^{i s z} \,,
\end{equation}
with $\eta_\omega \in \mathbb{R}$. Equation~\eqref{diracnltife} yields the conditions
\beq
s&=&\frac{1}{\eta_\omega}\left[\omega -q \Phi_0-\mu+\lambda_0\, |T_+|^2(1-\eta_\omega^2) \right]\,,\\
\eta_\omega^2&=&\frac{q \Phi_0-\omega-\mu}{2 \lambda_0 |T_+|^2} \nonumber\\
&\pm& \sqrt{\left(\frac{q\Phi_0-\omega-\mu}{2 \lambda_0 |T_+|^2}-1\right)^2-\frac{2 \mu}{\lambda_0 |T_+|^2}}\,. \label{etaks}
\eeq
Since we are looking for solutions with real $\eta_\omega$, we focus on modes such that
\beq \label{range}
\mu<\omega<q\Phi_0-\mu-2\lambda_0 |T_+|^2-\sqrt{8 \mu \lambda_0 |T_+|^2} \, .
\eeq
Using the $z$-component of the particle-number current,
\begin{equation} 
j^{z}=\frac{1}{2}\bar{\Psi}\gamma^{3} \Psi \,,
\end{equation}
one can show that 
\be
\label{current}
\left|\frac{(j^r)^z}{(j^i)^z}\right|= 1- \eta_\omega \left(\frac{\omega+\mu}{k}\right)  \frac{|T_+|^2}{|I_+|^2} \,.
\ee
It is easy to see that superradiant solutions must have $\eta_\omega<0$. We have two possible choices for a negative $\eta_\omega$ in \eqref{etaks}.

Assuming that the notion of group velocity is preserved in the non-linear case (at least for a very small $\lambda_0$), we need to make sure that the solutions satisfy the correct boundary conditions (\textit{i.e.}, the asymptotic solution \eqref{transnl} must correspond to a transmitted wave).
First, let us consider the $\eta_\omega^-<0$ satisfying
\beq \nonumber
(\eta_\omega^-)^2&=&\frac{q \Phi_0-\omega-\mu}{2 \lambda_0 |T_+|^2}- \sqrt{\left(\frac{q\Phi_0-\omega-\mu}{2 \lambda_0 |T_+|^2}-1\right)^2-\frac{2 \mu}{\lambda_0 |T_+|^2}}\,. 
\eeq
Interestingly, one can show that this solution also exists for $\lambda_0=0$ (\textit{i.e.} the limit $\lambda_0 \to 0$ is finite). Moreover, the wave $e^{-i(\omega t - s^- z)}$ has group velocity
\begin{align}\nonumber
	v_g^-=\left(\frac{d s^-}{d \omega}\right)^{-1}=-\left(\frac{d\eta_\omega^-}{d \omega}\right)^{-1}\frac{(\eta_\omega^-)^2}{s \eta_\omega^-+\lambda_0|T_+|^2[1+(\eta_\omega^-)^2]}\,.
\end{align}
 By doing a numerical analysis, one can conclude that $v_g^-<0$. This asymptotic solution does not satisfy the correct boundary conditions, since its group velocity is not compatible with a transmitted wave. Then, we exclude this solution.
 On the other hand, the solution $\eta_\omega^+<0$ satisfying
\beq \nonumber
(\eta_\omega^+)^2&=&\frac{q \Phi_0-\omega-\mu}{2 \lambda_0 |T_+|^2}+ \sqrt{\left(\frac{q\Phi_0-\omega-\mu}{2 \lambda_0 |T_+|^2}-1\right)^2-\frac{2 \mu}{\lambda_0 |T_+|^2}}\,. 
\eeq
exists only for $\lambda_0>0$, and the wave $e^{-i(\omega t - s^+ z)}$ has group velocity
\begin{align}\nonumber
v_g^+=\left(\frac{d s^+}{d \omega}\right)^{-1}=-\left(\frac{d\eta_\omega^+}{d \omega}\right)^{-1}\frac{(\eta_\omega^+)^2}{s \eta_\omega^++\lambda_0|T_+|^2[1+(\eta_\omega^+)^2]}\,.
\end{align} 
A numerical inspection allows us to conclude that $v_g^+>0$ for modes $\omega$ in a subset of \eqref{range}. These modes satisfy the appropriate boundary conditions, and such a solution is interpreted as a transmitted wave. Then, this non-linear fermionic theory admits solutions describing superradiant scattering on this setup. It can be shown that the mass plays a key role in the existence of these solutions. Notice that despite the fact that superradiant scattering is not possible in the $\lambda_0=0$ case, one can attain total reflection if the fermions are massive.

Additionally, we solved numerically Eq.~\eqref{diracnltife} with boundary condition \eqref{transnl}, 
for modes $\omega$ such that $v_g^+>0$. Our results indicate that superradiance is indeed present. Thus, the numerical results are consistent with the analytical ones: both indicate the presence of superradiance in this setup.

One may notice that the coupling that we considered breaks the gauge invariance of the theory, since the term $A_\mu A^\mu$ is not invariant under gauge transformations. This fact, however, is not relevant for our discussion, since we are not concerned about the generality or physical validity of this theory. It is interesting that the theory can be made gauge invariant if the number of spatial dimensions is larger than $1$ (in particular, in the physically relevant case of $3$ spatial dimensions). This has to do with the fact that, in these setups, one can show that $\bar{\Psi} \Psi$ vanishes naturally at infinity (to conserve the flux of the Noether's current), and, thus, the coupling with $A_\mu A^\mu$ is not necessary. Remember that the reason to consider this kind of coupling was to make Eq.~\eqref{diracfe} linear at infinity. We expect our results to be extended easily to these physically relevant setups.


\section{Superradiance in black hole backgrounds}
It is well known that static, charged BHs are described by the so-called Reissner-Nordstr\"{o}m (RN) geometry. In spherical coordinates, the RN geometry is represented by the squared line element
\begin{equation} \label{RNmetric}
ds^2=f \, dt^2- f^{-1} \, dr^2 - r^2(\, d\theta^2 + \sin^2 \theta \, d\varphi^2) \,.
\end{equation}
Here,
\begin{equation}\label{f}
f(r)=1-\frac{2 M}{r}+\frac{Q^2}{r^2} \,,
\end{equation}
where $M$ and $Q$ are the mass and electric charge of the BH, respectively. 
In these coordinates, there is an event horizon at
\begin{equation}
r=r_+=M+\sqrt{M^2-Q^2} \,.
\end{equation}
The charge $Q$ sources a spherically symmetric field
\begin{equation} \label{em_covector}
A_\mu= \left(\Phi(r), \vec{0}\right) \quad \quad \text{with} \quad \quad \Phi(r)=\frac{Q}{r} \,.
\end{equation}
As in the case of the Klein paradox, we use the test field approximation, ignoring the back-reaction on the geometry of spacetime. Although we use charged fields, we also ignore the electromagnetic field produced by them. These approximations are justified by the fact that these effects are of second order on the charged fields and for sufficiently weak (small amplitude) fields, can be neglected. Moreover, in astrophysical relevant setups, the electromagnetic field produced by such charged fields have negligible effect on the geometry~\cite{Mosta:2009rr}.
\subsection{Scattering of scalar fields}
Consider the scalar field theory
\begin{equation}
S_{\rm scalar}= \int dx^4 \sqrt{-g} \left[\,g_{\mu \nu} D^\nu\phi(D^\mu\phi)^*-\mu^2|\phi|^2\right] \,,
\end{equation}
with $$D_\mu= \nabla_\mu+ i q A_\mu \quad,$$ and all the other quantities defined as in the Klein setup case. 
From this action, we obtain the equation of motion
\begin{equation} \label{scalarfec}
	D_\mu D^\mu \phi + \mu^2 \phi=0 \,.
\end{equation}
Using the ansatz
\begin{equation} \label{ansatzscalar}
\phi_\omega (t, r, \theta, \varphi)=\sum_{l, m} e^{-i\omega t}\, Y_l^{m}(\theta, \varphi) \frac{\psi_\omega(r)}{r} \,,
\end{equation} 
we obtain 
\beq 
\label{scalartifec}
&&f^2 \frac{d^2}{d r^2} \psi_\omega + f f'\frac{d}{d r} \psi_\omega+\left[\left(\omega -q \Phi\right)^2-V\right] \psi_\omega=0\,,\\
&&V=f\left(\frac{l(l+1)}{r^2}+\frac{f'}{r} + \mu^2 \right) \,.
\eeq
It is possible to show that this equation admits the asymptotic solution 
\begin{equation}
\psi_\omega(r \to +\infty)=I \, e^{-i  k r}+ R \, e^{+i k r} \,,
\end{equation} 
with $k$ satisfying Eq.~\eqref{drldks}. 
The transmitted wave can be written as
\begin{align}\label{transscalar}
\begin{split}
\psi_\omega(r \to r_+)=&T\, \exp\big[-i s \frac{r_+^2}{r_+-r_-} \log(r-r_+)\big] \,,
\end{split}
\end{align}
where 
\begin{equation} \label{drsc}
s= \omega-q \Phi_+ \,,
\end{equation}
with $\Phi_+=\Phi(r_+)$. The sign of $s$ is such that the wave $e^{-i(\omega t -s r_*)}$ has group velocity
\begin{equation}\nonumber
	v_g=\left(\frac{d s}{d \omega}\right)^{-1}=1>0\,,
\end{equation}
where we used the so-called tortoise coordinate
\begin{equation} \nonumber
	r_*=\frac{r_+^2}{r_+-r_-}\text{log}(r-r_+)\,.
\end{equation}
Then, the wave $e^{-i(\omega t +s r_*)}$ has negative $v_g$ (\textit{i.e.} moves towards the event horizon), which means that $s$ has the correct sign for~Eq.\eqref{transscalar} to be interpreted as a transmitted wave. 
The quantities $(R/I)$ and $(T/I)$ appearing in the above solutions are complex functions of $l$, $m$ and $\omega$ (see Eq.~\eqref{ansatzscalar}) but, for the sake of simplicity, we omit this dependence in our notation. 

Using the particle-number current
\begin{equation}
j^\mu=- \frac{i}{2}\,[ \phi^* D^\mu \phi - \phi\, (D^\mu \phi)^*] \,, 
\end{equation}
and its flux $\mathcal{F}$ over a spherical surface of radius $r$ (denoted by $S_r$) with $r \to +\infty$,
\begin{equation} \label{scalarfluxc}
\mathcal{F}=\lim_{r \to +\infty} \int_{S_r} d\Omega\, r^2 j^r \,,
\end{equation} 
one can show
\begin{equation}
\left| \frac{\mathcal{F}^r}{\mathcal{F}^i} \right|=1-\frac{s}{k} \frac{\sum_{l, m} |T|^2}{\sum_{l, m} |I|^2} \,,
\end{equation}
where $\mathcal{F}^i$ and $\mathcal{F}^r$ are the fluxes of the incident and reflected currents, respectively.
By definition, there is superradiant scattering when $|\mathcal{F}^r|>|\mathcal{F}^i|$. Thus, the scalar field has the superradiant modes 
\begin{equation}
\mu<\omega<q \Phi_+ \,.
\end{equation}
These modes are equal to the ones obtained in Refs.~\cite{Brito:2015oca,Bekenstein:1973mi}.
\subsection{Scattering of Dirac fields}
Here, we consider the curved spacetime Dirac theory
\begin{equation}
S_\text{Dirac}=\int dx^4 \sqrt{-g} \,[\,i \, \bar{\Psi} G^\mu \mathcal{\widetilde{D}}_\mu \Psi-\mu \bar{\Psi} \Psi\,] \,,
\end{equation}
with $\mu$ the mass of the Dirac field, $G^\mu$ the Dirac matrices in curved spacetime and $\mathcal{\widetilde{D}}_\mu=\partial_\mu+i q A_\mu -\Gamma_\mu$, where $q>0$ is the electric charge of the field and $\Gamma_\mu$ is the so-called spin connection.
We use the following tetrad,
\begin{align}
\begin{split}
e_a^{\,\;t}&=\left(\,\frac{1}{\sqrt{f}},\, 0,\, 0,\, 0\,\right) \,,\\
e_a^{\,\;r}&=\left(\,0,\, \sqrt{f} \sin \theta \cos \varphi,\, \sqrt{f}\sin \theta \sin \varphi ,\, \sqrt{f} \cos \theta \,\right) \,,\\
e_a^{\,\;\theta}&=\left(\,0,\, \frac{1}{r} \cos \theta \cos \varphi,\, \frac{1}{r} \cos \theta \sin \varphi ,\, -\frac{\sin \theta}{r} \,\right)\,, \\
e_a^{\,\;\varphi}&=\left(\,0,\, -\frac{1}{r} \frac{\sin \varphi}{\sin \theta} ,\, \frac{1}{r} \frac{\cos \varphi}{\sin \theta} ,\, 0\,\right) \,.
\end{split}
\end{align}
As shown in Ref.~\cite{Yau2}, with such tetrad the field equation reads
\beq
&&\gamma^0\left( \frac{i}{S(r)} \frac{\partial}{\partial t}-\frac{q}{S(r)} \Phi(r) \right)\Psi
+\left(i \gamma^\theta \frac{\partial}{\partial \theta}+i \gamma^\varphi \frac{\partial}{\partial \varphi} -\mu \right) \Psi\nonumber\\
&&+ \gamma^r \left(i S(r)\frac{\partial}{\partial r} +\frac{i}{r}(S(r)-1)+\frac{i}{2} S'(r) \right)\Psi=0\,,
\label{diracRN}
\eeq
where 
\be
S(r)=\sqrt{f(r)}\,,
\ee
and $S'(r)$ is the radial derivative of $S$. The matrix functions $\gamma^r$, $\gamma^\theta$ and $\gamma^\varphi$ appearing in the above equation are defined by
\begin{align}
\begin{split}
\gamma^r&= \gamma^1 \sin \theta \, \cos \varphi + \gamma^2 \sin \theta \, \sin \varphi+\gamma^3 \cos \theta \,,\\
\gamma^\theta&= \frac{1}{r} \left(\gamma^1 \cos \theta \, \cos \varphi+\gamma^2 \cos \theta \, \sin \varphi -\gamma^3 \sin \theta\right) \,,\\
\gamma^\varphi&=  \frac{1}{r \sin \theta} \left( -\gamma^1 \sin \varphi + \gamma^2 \cos \varphi \right) \,.
\end{split} 
\end{align}

Let us consider the following ansatz for the Dirac spinor:
\beq
&&\Psi_{j\,k\,\omega}^\pm(t,r,\theta, \varphi) = e^{-i \omega t} \frac{S^{-1/2}}{r} \nonumber \\ 
&&\times \left( \begin{array}{c}
\chi_{j\mp 1/2}^{k}(\theta, \varphi) \,\, \Upsilon_{j\,k\,\omega\,1}^\pm(r)  \\
i \, \chi_{j\pm 1/2}^{k}(\theta, \varphi) \,\, \Upsilon_{j\,k\,\omega\,2}^\pm(r)  \end{array} \right)\,,
\eeq
with the two-spinors $\Upsilon_{j \, k\, \omega}^\pm$. Here, the $\chi_{j-1/2}^{k}$ and $\chi_{j+1/2}^{k}$ are the so-called spinor spherical harmonics ($j$ and $k$ are half-integers satisfying $j\geq|k|$)~\cite{Biedenharn_Louck_1984}. 

Using the ansatz above, after a little algebra we obtain the matrix equations
\begin{widetext}
\be
S \frac{d}{d r}\Upsilon_{j\,k\,\,\omega}^\pm=\left[ \frac{\omega-q \Phi}{S}
\left( \begin{array}{cc}
0 & -1  \\
1 & 0  \end{array} \right) \pm \frac{2 j +1}{2 r}
\left( \begin{array}{cc}
1 & 0  \\
0 & -1  \end{array} \right)-\mu
\left( \begin{array}{cc}
0 & 1  \\
1 & 0  \end{array} \right)
\right] \Upsilon_{j\,k\,\,\omega}^\pm \,.\label{mastereq}
\ee
\end{widetext}
These equations are in accordance with those obtained in Ref.~\cite{Finster:1998ak}. 
Their asymptotic solutions are 
\beq
&&\Upsilon_{j\,k\,\,\omega}^\pm(r\to+ \infty)\nonumber\\
&&=I^\pm e^{-i k r} \left( \begin{array}{c}
	1 \\
	i \sqrt{\dfrac{\omega-\mu}{\omega+\mu}} \end{array} \right)\nonumber + R^\pm e^{i k r} \left( \begin{array}{c}
	1 \\
	-i \sqrt{\dfrac{\omega-\mu}{\omega+\mu}} \end{array} \right)\,, 
\eeq
with $k$ satisfying Eq.~\eqref{drldks}, and 
\begin{align}
\begin{split}
\Upsilon_{j\,k\,\,\omega}^\pm(r\to r_+)=T^\pm e^{-i s r_* }  \left( \begin{array}{c}
	1 \\
	i  \end{array} \right) \,,
	\end{split}
\end{align}
with $s$ satisfying Eq.~\eqref{drsc}.
Both $(R^+/I^+)$, $(R^-/I^-)$, $(T^+/I^+)$ and $(T^-/I^-)$ are complex-valued functions of $j$, $k$ and $\omega$.

A solution $\Psi_\omega$ of the Dirac equation can always be expressed in the form
\begin{equation} \label{diracgen}
\Psi_\omega=\sum_{j, \, k} \left(\Psi_{j\,k\,\omega}^+ +\Psi_{j\,k\,\omega}^- \right) \,.
\end{equation}
This is because, by the definition of spinor spherical harmonics, with an expression of the form \eqref{diracgen} one can construct every combination of spherical harmonics in each one of the four components of the spinor.

Using the particle-number current
\begin{equation} \label{pnumbercur}
J^\mu=\frac{1}{2}\bar{\Psi}G^\mu \Psi \,,
\end{equation}
one can show 
\begin{equation}
\left|\frac{\mathcal{F}^r}{\mathcal{F}^i}\right|=1-\sqrt{\frac{\omega+\mu}{\omega-\mu}} \, \frac{|T^+|^2+|T^-|^2}{|I^+|^2+|I^-|^2} \,.
\end{equation}
Since $|\mathcal{F}^r| \leq |\mathcal{F}^i|$, the Dirac field does not have any superradiant mode on this background. This result was known as a limit of the more general Kerr-Newman geometry \cite{Lee:1977gk}. 

\subsection{Scattering of non-linear Dirac fields}
Now we would like to show that the non-linear interactions considered in the Klein setup do not allow Dirac fields around BHs to be superradiantly amplified. This is in contrast  with what we obtained on flat spacetime, where amplification was allowed. Consider again the non-linear Dirac theory \cite{SOLER},
\beq \label{Diracactionnc}
S=\int dx^4 \sqrt{-g} \left(\,i \, \bar{\Psi} G^\mu \mathcal{\widetilde{D}}_\mu \Psi-\mu \bar{\Psi} \Psi\,+\frac{\lambda}{2}(\bar{\Psi}\Psi)^2\right)\nonumber \,,
\eeq
with the coupling
\begin{equation} 
\label{couplingnl}
\lambda(r)=\widetilde{\lambda}\, q^2 A_\mu A^\mu=
\widetilde{\lambda}\, q^2 \,\frac{Q^2}{r^2} >0 \,,
\end{equation}
where $\widetilde{\lambda}>0$ is a real constant. In the same tetrad of the last section, the field equation associated with this action is
\beq
&&\gamma^0 \left( \frac{i}{S(r)} \frac{\partial}{\partial t}-\frac{q}{S(r)} \Phi(r) \right)\Psi\nonumber\\
&&+\gamma^r \left(i S(r)\frac{\partial}{\partial r} +\frac{i}{r}(S(r)-1)+\frac{i}{2} S'(r) \right)\Psi\nonumber\\
&&+\left(i \gamma^\theta \frac{\partial}{\partial \theta}+i \gamma^\varphi \frac{\partial}{\partial \varphi} -\mu + \lambda \bar{\Psi} \Psi \right) \Psi=0 \,.\label{diraceqRNnl}
\eeq

Let us consider the particular ansatz
\be
\Psi_{\omega}(t,r,\theta, \varphi) =N e^{-i \omega t} \frac{S^{-1/2}}{r} \left( \begin{array}{c}
\chi_{j-1/2}^{k}(\theta, \varphi) \,\, F(r)  \\
i \, \chi_{j+1/2}^{k}(\theta, \varphi) \,\, G(r)  \end{array} \right) \nonumber\,,
\ee
with $j=k=+\frac{1}{2}$. Using this ansatz, after a little algebra one gets
\beq
&&-SF'+\frac{F}{r}\nonumber \\
&=&\left( \frac{\omega-q \Phi}{S}+\mu-\frac{\lambda}{4\pi r^2 S} |N^2| \,\big(|F|^2-|G|^2\big) \right)G \,, \label{diracnltifecs1}\\
&&SG'+\frac{G}{r}\nonumber\\
&=&\left( \frac{\omega-q \Phi}{S}-\mu+\frac{\lambda}{4\pi r^2 S} |N^2|\,\big(|F|^2-|G|^2\big) \right)F\label{diracnltifecs2}	\,,
\eeq
where $'$ means derivative with respect to $r$.
These equations admit solutions with the asymptotics
\beq
	&&\left( \begin{array}{c}
	 F  \\
	 G  \end{array} \right)(r \to +\infty) \nonumber\\ 
	&=&I^+ e^{-i k r} \left( \begin{array}{c} \nonumber
	 1 \\
	 i \sqrt{\dfrac{\omega-\mu}{\omega+\mu}} \end{array} \right)+ R^+ e^{i k r} \left( \begin{array}{c}
	 1 \\
	 -i \sqrt{\dfrac{\omega-\mu}{\omega+\mu}} \end{array} \right)\,,
\eeq
where $k$ satisfies Eq.~\eqref{drldks}. Building on the previous section, let us search for solutions with the asymptotics 
\begin{align}\label{transnlcs}
\begin{split}
&\left( \begin{array}{c}
F  \\
G  \end{array} \right)(r \to r_+)= T^+ \left( \begin{array}{c}
1 \\
i\, \eta_\omega \end{array} \right) e^{-i s r_*} \,,
\end{split}
\end{align}
with $r_*$ defined as in that section and $\eta_\omega \in \mathbb{R}$. 
After some algebra, Eqs.~\eqref{diracnltifecs1}-\eqref{diracnltifecs2} give us the conditions
\begin{align}
	s&=\frac{1}{\eta_\omega}\left[\omega-q\Phi_+ + \lambda_+|T^+|^2 (1-\eta_\omega^2)\right] \,,\\
	s^2&=(\omega-q\Phi_+)^2-\lambda_+^2\, |T^+|^4(1-\eta_\omega^2)^2\,,
\end{align}
with $\lambda_+$ defined by
\beq
\lambda_+=\frac{\lambda(r_+)}{4 \pi r_+^2} \nonumber \,.
\eeq
The above two equalities imply that
\begin{equation}
\begin{split}
(1-\eta_\omega^2)\big[\left(\omega-q\Phi_+\right)^2 +2 (\omega&-q\Phi_+) \lambda_+ |T^+|^2\\ &+\left(1-\eta_\omega^4\right) \lambda_+^2 |T^+|^4\big]=0 \,.
\end{split}
\end{equation}
There are four possible choices for $\eta_\omega$:
\begin{equation}\label{eta1}
	\eta_\omega=\pm 1 \,,
\end{equation}
\begin{equation}\label{eta2}
	\eta_\omega=\pm \left(1+\frac{q\Phi_+-\omega}{\lambda_+^2 |T^+|^4}\left(1-2 \lambda_+ |T^+|^2\right)\right)^{1/4}\,,
\end{equation}
where in the last expression the modes $\omega$ must be such that the expression is real. Using the particle-number current
\begin{equation} 
J^\mu=\frac{1}{2}\bar{\Psi}G^\mu \Psi \,,
\end{equation}
one gets
\begin{equation}
\left|\frac{\mathcal{F}^r}{\mathcal{F}^i} \right| =1-\eta_\omega\, \sqrt{\frac{\omega+\mu}{\omega-\mu}} \frac{|T^+|^2}{|I^+|^2} \,. 
\end{equation}
We see that superradiant solutions must have negative $\eta_\omega$. Thus, we have two possible choices for $\eta_\omega$. Moreover, the solutions must satisfy the correct boundary conditions (\textit{i.e.} $\eta_\omega$ must be such that \eqref{transnlcs} corresponds to a transmitted wave). 

Let us start with the solution $\eta_\omega^-=-1$. Notice that this solution also exists for $\lambda_+=0$. Furthermore, the wave $e^{-i(\omega t - s^- r_*)}$ has group velocity $$v_g^-=\left(\frac{d s^-}{d \omega}\right)^{-1}=-1\,.$$ Then, the wave $e^{-i(\omega t + s^- r_*)}$ has positive $v_g$ (\textit{i.e.}, travels away from the BH), and the solution does not satisfy the correct boundary conditions (it is not a transmitted wave). So, we exclude this solution. The other solution $$\eta_\omega^+=- \left(1+\frac{q\Phi_+-\omega}{\lambda_+^2 |T^+|^4}\left(1-2 \lambda_+ |T^+|^2\right)\right)^{1/4}\,,$$ exists only for $\lambda_+>0$. Moreover, the wave $e^{-i(\omega t - s^+ r_*)}$ has group velocity $$v_g^+=\left(\frac{d s^+}{d \omega}\right)^{-1}=-\left(\frac{d\eta_\omega^+}{d \omega}\right)^{-1}\frac{(\eta_\omega^+)^2}{s \eta_\omega^++\lambda_+|T^+|^2[1+(\eta_\omega^+)^2]}\,.$$ 
Again, we conclude that $v_g^+$ is negative for all modes $\omega$. Then, the wave $e^{-i(\omega t + s^+ r_*)}$ has positive group velocity. So, this solution does not satisfy the correct boundary conditions, and it is also excluded. We conclude that, in this setup, there is no amplified solution satisfying the correct boundary conditions.
Then, although this non-linear fermionic theory admits superradiant scattering solutions in flat spacetime, it does not admit them in BH spacetimes. The reason for this is that, in these kind of backgrounds, the mass term of the scattering fields vanishes at the horizon. And, as we stated in the Klein setup treatment, it is the mass term which makes possible the existence of superradiant scattering solutions.

Interestingly, the fact that the mass term vanishes at the horizon favours superradiant scattering of bosonic (scalar) fields. Notice that while in the Klein setup the necessary condition for the existence of superradiant scattering modes is $qV_0>2 \mu$; on RN this condition is $qV_+>\mu$.

We should remark that finding a non-linear Dirac theory with solutions describing superradiant scattering does not seem to be a trivial task. Besides the non-linear theory \eqref{Diracactionnc} provided above, we analyzed the theories
\begin{equation}
	S=\int dx^4 \sqrt{-g} \left(\,i \, \bar{\Psi} G^\mu \mathcal{\widetilde{D}}_\mu \Psi+i \lambda(\bar{\Psi}\Psi) \bar{\Psi} G^\mu \mathcal{\widetilde{D}}_\mu \Psi \right)\nonumber \,, 
\end{equation}
\begin{equation}
		S=\int dx^4 \sqrt{-g} \left(\,i \, \bar{\Psi} G^\mu \mathcal{\widetilde{D}}_\mu \Psi- \frac{\lambda}{2}(\overline{\bar{\Psi} G^\mu \mathcal{\widetilde{D}}_\mu \Psi}) \bar{\Psi} G^\mu \mathcal{\widetilde{D}}_\mu \Psi \right)\nonumber \,,
\end{equation}
with $\lambda$ as in \eqref{Diracactionnc}, finding no superradiant solutions (neither on RN spacetime nor on Klein setup).

\section{Superradiant amplification of wavepackets}

Up to now, we used a standard Fourier-transform approach to study the scattering of monochromatic (delocalized) solutions. More realistic solutions, describing
compact initial data, should instead focus on time-domain analysis. We now turn to this problem; to construct a pulse-type solution as initial condition, we use a packet of monochromatic waves. 
Our purpose is two-fold: we want to show that the amplification factors obtained using this approach are consistent with the ones obtained with monochromatic waves. In addition, we will show 
that amplification occurs even when no horizon is present, contrary to what some studies (in Fourier-space) suggest. Our conclusions agree, of course, with the expectation that a spacetime region which is causally disconnected from some phenomenon can have no bearing in its evolution. Ergoregions are the sole responsible for superradiance.
The absence of horizons merely disguises superradiant amplification, by triggering an ergoregion instability. We show how both effects are compatible and seen simultaneously.

\subsection{Fourier-domain calculations\label{sec:fourier}}
Consider again the scattering of a (for simplicity, massless) scalar field on the Klein setup. This problem admits the monochromatic solutions
\begin{equation} \label{phiwp}
\phi_\omega(t, z)=e^{-i \omega t} \varphi_\omega(z) \,,
\end{equation}
where $\varphi_\omega$ satisfies Eq.~\eqref{scalartife} and has the asymptotics given by Eqs.~\eqref{scalariI} and \eqref{scalariII} (here, with $\mu=0$). Thus, the wave packet solution has the form
\begin{equation} \label{phiwp1}
	\phi(t,z)=\int d\omega\, e^{-i \omega t} \varphi_\omega(z)  \,.
\end{equation} 
To proceed with an analytic study of this problem, we focus on the simplest of the potential barriers: a step potential 
\be
q \Phi(z)= q\Phi_0 \mathcal{H}(z)\,,\label{heaviside_potential}
\ee
with $\mathcal{H}(z)$ the Heaviside step function. With this barrier, it is trivial to show (using continuity of $\varphi_\omega(z)$ and its first derivative at $z=0$) that the amplitudes of the monochromatic waves satisfy
\begin{equation} \label{stepcoef}
\frac{R}{I}=\frac{q \Phi_0}{2 \omega-q \Phi_0}\,,\qquad
\frac{T}{I}=\frac{2 \omega}{2 \omega- q \Phi_0} \,.
\end{equation}
Moreover, since Eq.~\eqref{scalartife} is linear, one has the freedom to fix $I$ as a function of $\omega$.

\begin{figure}[H]
\centering
\includegraphics[width=1\linewidth]{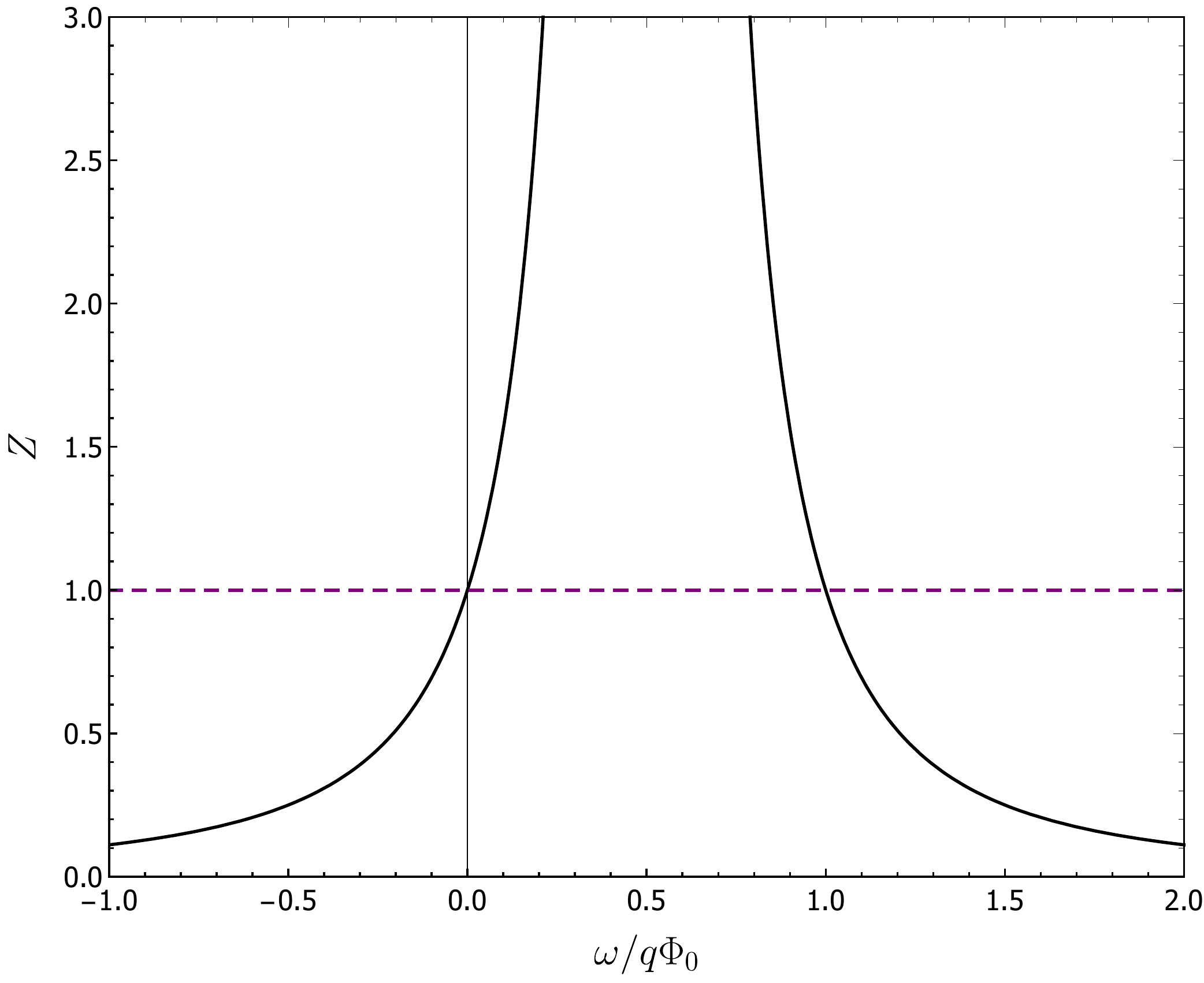}
\caption{Amplification factor for a monochromatic scalar wave of frequency $\omega$ scattered off a step potential~\eqref{heaviside_potential}. 
}
\label{fig:ZStepPot}
\end{figure}

Using Eq.~\eqref{scalaramp} one obtains the amplification factor
\begin{equation}
Z=\left|\frac{j_z^r}{j_z^i}\right|=\left(\frac{q \Phi_0}{2 \omega- q \Phi_0}\right)^2 \,.
\end{equation}
As it is seen in the above expression, the amplification factor diverges for $\omega= q\Phi_0/2$. This happens because the system has one (real) normal mode with $\omega=q\Phi_0/2$. The form of this amplification factor is shown in Fig.~\ref{fig:ZStepPot}.
\begin{figure*}[ht]
\begin{tabular}{cc}
\includegraphics[width=0.45\textwidth,clip]{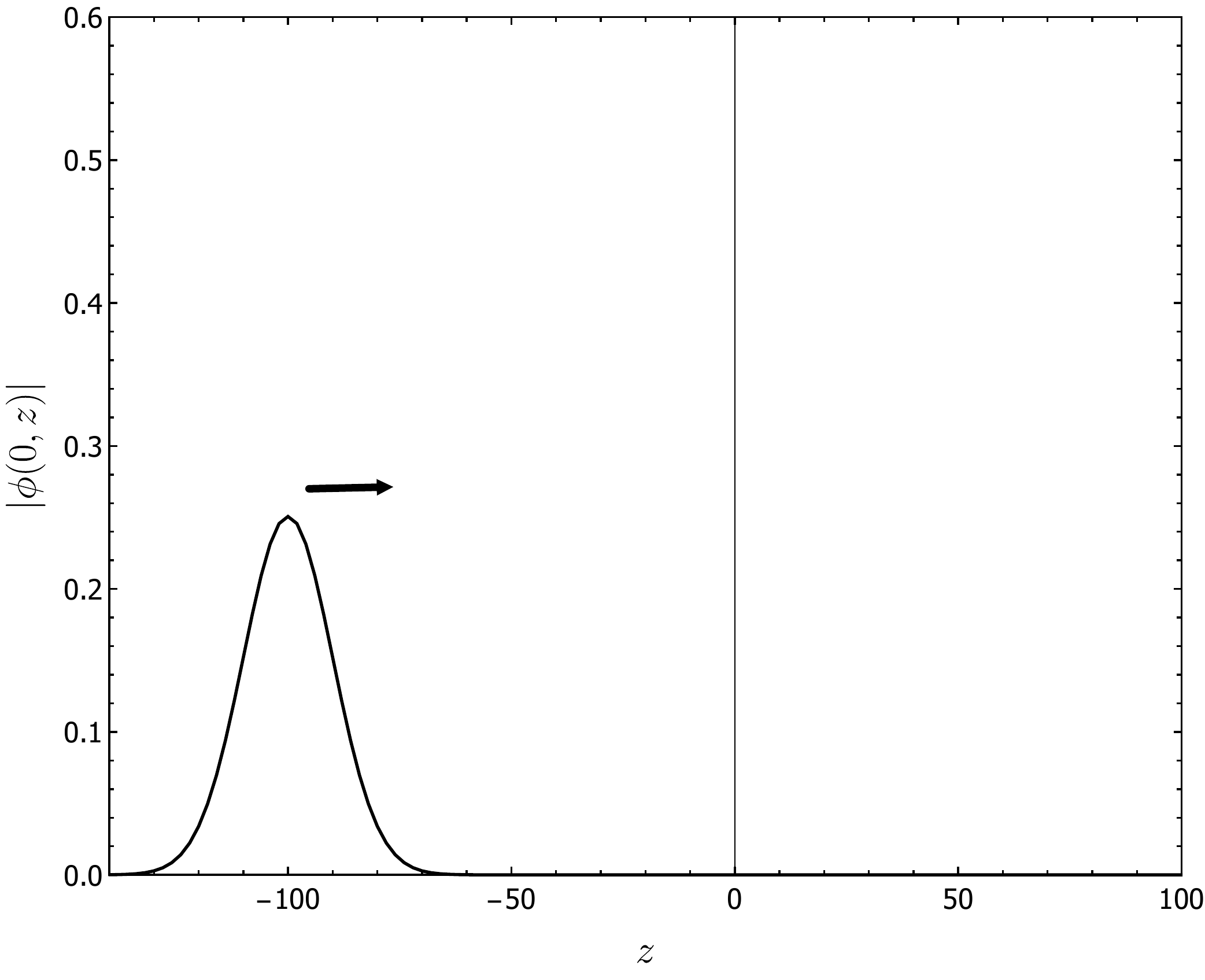}&
\includegraphics[width=0.45\textwidth,clip]{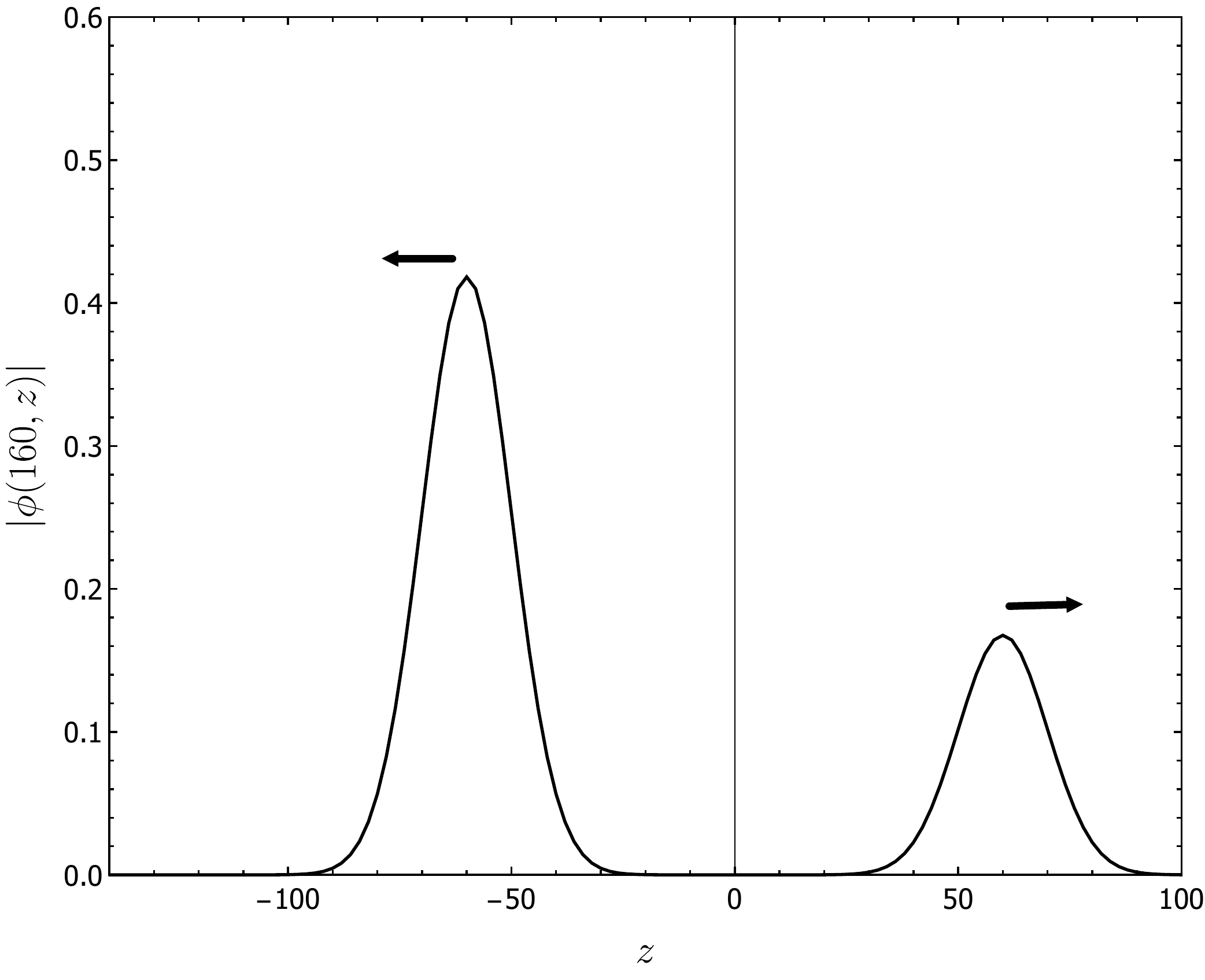}
\end{tabular}
\caption{Wave packet solution at $t=0,\, 160$ (left and right panel, respectively) for a Gaussian wavepacket scattered off a step-potential~\eqref{heaviside_potential}, characterized by $q\Phi_0=10$. The initial data corresponds to a Gaussian centered at $\omega_0=2$, with $\sigma_\omega=0.1$, $z_0=-100$. 
At $t=160$ the initial pulse scattered off the barrier, a fraction was reflected and parts penetrated the barrier. It is apparent that the process is superradiant: the right-moving pulse at $t=0$ was converted into a higher-amplitude, left-moving pulse at t=160.
As a reference, for a perfectly monochromatic pulse with frequency $\omega=2$ incident on such a potential barrier, the amplification factor is $Z\simeq2.778$.}
\label{fig:phi0}
\end{figure*}
Suppose, instead, one wants to understand how a wave packet is scattered. To be concrete, consider a Gaussian distribution in frequencies centered at $\omega_0$, with standard deviation $\sigma_\omega$, and such that at instant $t=0$ it is a pulse centered at $z_0 \ll 0$. To construct this solution we use the freedom to fix $I(\omega)$. One can show that 
\begin{align}
I(\omega)&= e^{-\frac{(\omega-\omega_0)^2}{2 \sigma_\omega^2}} e^{i\omega z_0}\,,\label{gaussian}
\end{align}
describes such a pulse. Thus,
\begin{align}
\begin{split}
\varphi_\omega(z<0)&=I(\omega)\left(e^{i k z}+\frac{q \Phi_0}{2 \omega-q \Phi_0} e^{-i k z}\right)\, , \\
\varphi_\omega(z>0)&=I(\omega)\left(\frac{2 \omega}{2 \omega- q \Phi_0} \right) e^{i k z}\, ,
\end{split}
\end{align}
where we used \eqref{stepcoef}.

Using Eq.~\eqref{phiwp1} we evolved in time the initial profile, for $q\Phi_0=10,\omega_0=2,\sigma_\omega=0.1$ and $z_0=-100$. Our results are summarized in Fig.~\ref{fig:phi0}, which shows two snapshots of the field $\phi$ at $t=0$ (before scattering) and $t=160$ (after scattering), respectively. An important question concerns the total amplification of such a pulse for finite width $\sigma_\omega$. Is it smaller or larger than that of a corresponding monochromatic pulse? To study the presence of superradiant amplification we only need to integrate the flux of the $z$-component of the particle-number current \eqref{pncurrentscalar} from $t=-\infty$ to $t=+\infty$ at some $\widetilde{z}<0$. So, 
\beq
&&\int_{-\infty}^{+\infty} dt\, j_z(t,\widetilde{z})= \int_{-\infty}^{+\infty} dt\, \text{Im}\left(\phi^*\partial_{z} \phi \right)\nonumber\\ 
&&=\text{Im}\left(\int d\omega\,d\widetilde{\omega}\, \varphi_{\widetilde{\omega}}^* \partial_{z} \varphi_\omega \int_{-\infty}^{+\infty} dt\, e^{-i(\omega-\widetilde{\omega})t}\right) \nonumber\\
&&= 2 \pi \int d\omega\, \omega |I|^2 \left(1-\left|\frac{R}{I}\right|^2\right) \,.
\label{zscalarfluxint}
\eeq
This last integral gives the number of incident particles minus the number of reflected particles. If there is superradiant amplification, the above integral must be negative, but the converse is not true. This is because the number of incident anti-particles ($\omega<0$) is represented by a negative number of incident particles. In other words, the scattering of anti-particles on this barrier (which is always superradiant-less if the incident anti-particles come from left) gives always negative values for the integral \eqref{zscalarfluxint}.

Finally, to determine the number of incident particles, we can integrate the particle-number density ($t$-component of the Noether's 4-current) from $z=-\infty$ to $z=+\infty$ at $t=0$. One finds,
\begin{equation}
\int_{-\infty}^{+\infty} dz\, \left[-j_t(0,z)\right]= \int_{-\infty}^{0} dz\, \text{Im}\left(\phi^*\partial_{t} \phi \right) \,, 
\end{equation}
where we made use of the fact that $\phi(0,z>0) \sim 0$.
This integral can be shown to satisfy the equality
\begin{equation}
\int_{-\infty}^{0} dz\, \text{Im}\left(\phi^*\partial_{t} \phi \right)=2 \pi \int d\omega\, \omega |I|^2 \,.
\end{equation}
\begin{figure}[H]
\centering
\includegraphics[width=1\linewidth]{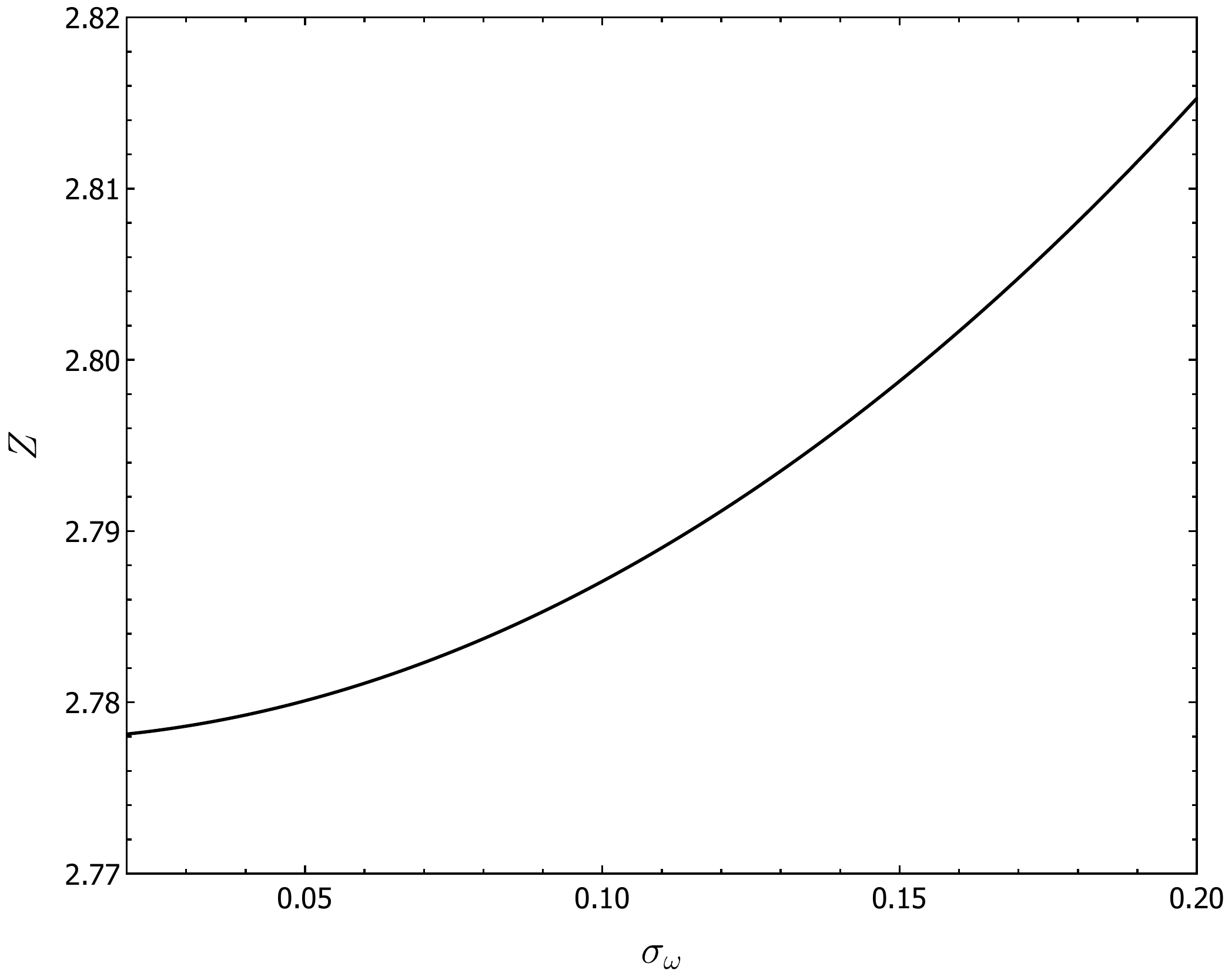}
\caption{Amplification factor for a Gaussian wavepacket described by \eqref{gaussian} as function of $\sigma_\omega$, for a coupling $q \Phi_0=10$ and $\omega_0=2$. For this scattering problem, monochromatic waves of frequency $\omega=2$ have $Z\simeq2.778$.}
\label{fig:ZScalarSigma}
\end{figure}
To evaluate the amplification factor of the scattering process, we need to evaluate the absolute value of the number of reflected particles divided by the absolute value of the number of incident particles: 
\be \label{Z}
Z=\left|\frac{\int_{-\infty}^{+\infty} dz\, \left[-j_t(0,z)\right]-\int_{-\infty}^{+\infty} dt\, j_z(t,\widetilde{z})}{\int_{-\infty}^{+\infty} dz\, \left[-j_t(0,z)\right]}\right|=\left|\frac{\int d\omega\, \omega |R|^2}{\int d\omega\, \omega |I|^2}\right|\,.
\ee
Our results are summarized in Fig.~\ref{fig:ZScalarSigma}, which refers to $\omega_0=2, q \Phi_0=10$, and shows the amplification factor $Z$ for different width $\sigma_\omega$. When the pulse is nearly monochromatic, i.e. at very small $\sigma_\omega$, the amplification asymptotes to the monochromatic-wave result ($Z=2.778$ at $\omega=2$, for those parameters). At larger width $\sigma_\omega$, the amplification turns out to be slightly larger, since the pulse now contains larger frequencies than $\omega=0.2\, q\Phi_0$, which are more prone to higher amplification in this particular potential barrier (cf. Fig.~\ref{fig:ZStepPot}).

\subsection{Time-domain analysis}

To complement our study of this type of scattering setups we have also performed numerical time evolutions of Eqs.~\eqref{scalarfe0},\eqref{scalarfec}.
We used as initial conditions,
\begin{equation} 
\label{ic1}
\phi(0, z)=\phi_0(z)=\exp\left[-\frac{(z-z_0)^2}{2 \sigma_z^2}\right] e^{i \omega_0 z}\,,
\end{equation}
and
\begin{equation}\label{ic2}
\partial_t \phi(0,z)+\partial_z \phi_0(z)=0 \,.
\end{equation}
The second initial condition is known as the advection equation and describes an initial pulse moving towards the potential barrier. We impose Sommerfeld outgoing conditions at the boundaries.
 For these numerical studies, we used two scattering setups: a charged scalar on a RN BH background and a Klein setup with potential
\begin{equation} 
\label{nearstep}
q\Phi(z)=\frac{q\Phi_0}{e^{-z/L}+1}\,,
\end{equation}
where $L\ll (2\pi)/\omega_0$.

\begin{figure*}[ht]
\begin{tabular}{ccc}
\includegraphics[width=0.45\textwidth,clip]{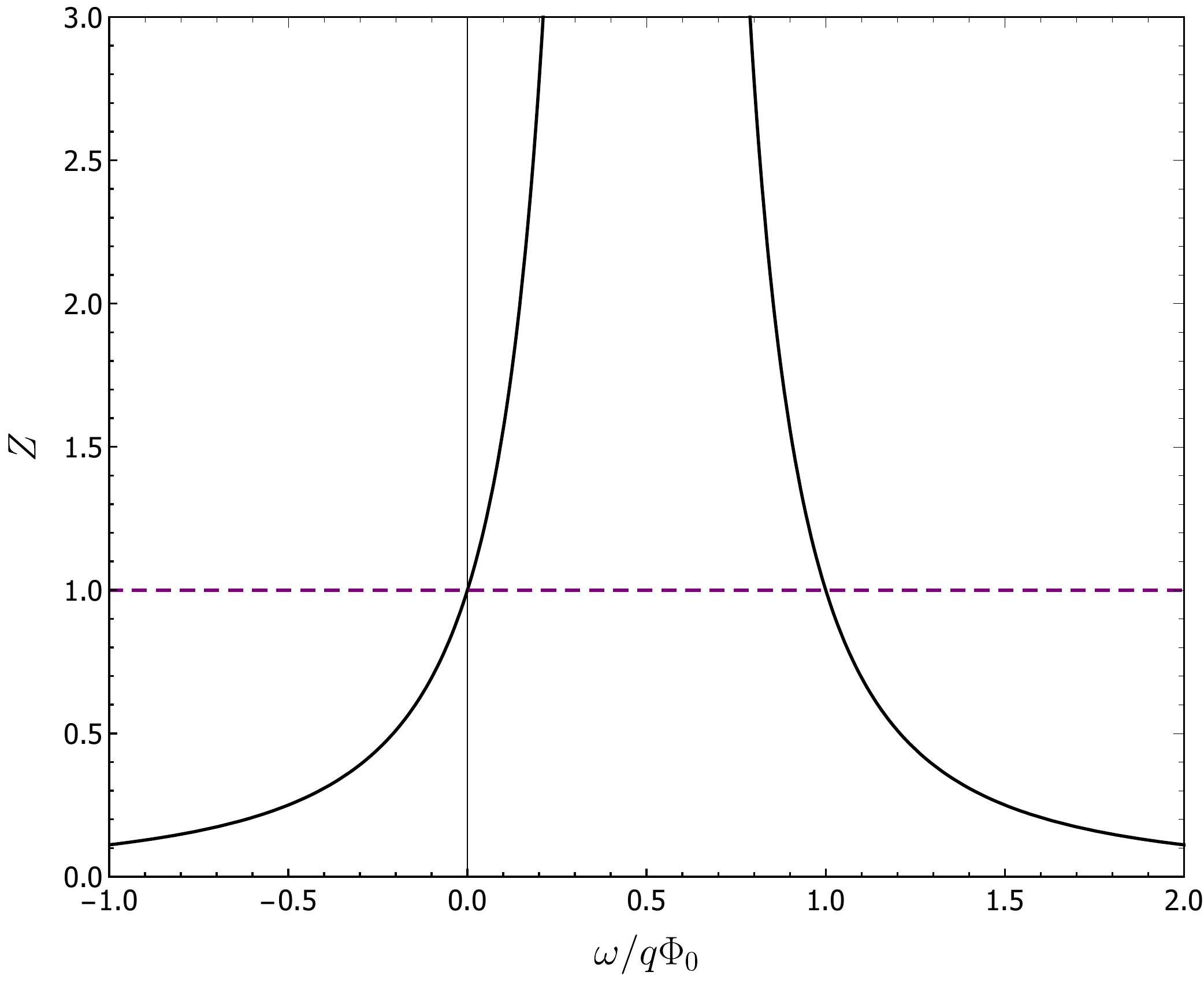}&
\includegraphics[width=0.45\textwidth,clip]{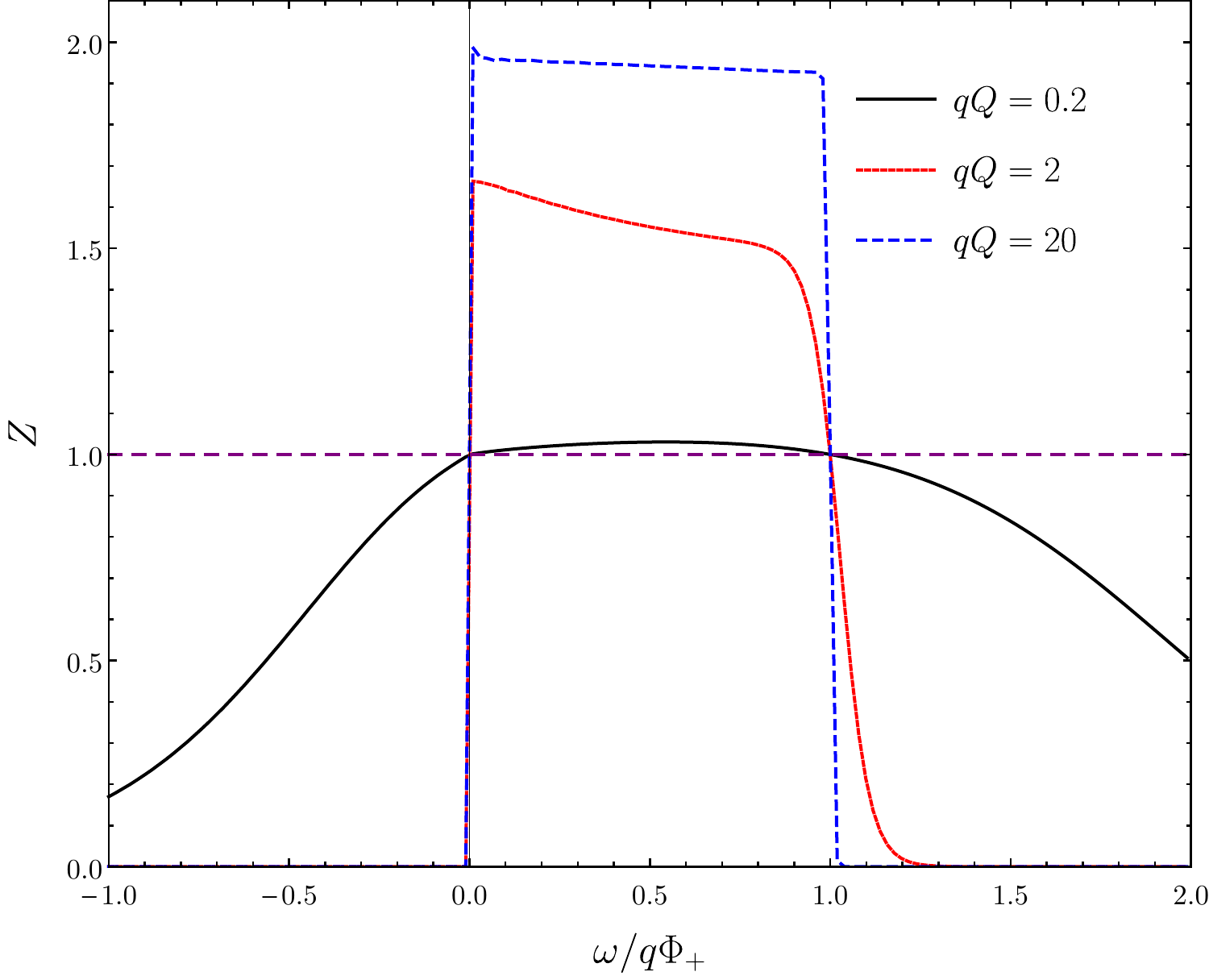}
\end{tabular}
\caption{Amplification factor for a monochromatic scalar wave scattering on the potential~\eqref{nearstep} (left, $L=0.001$) and
on a RN BH (right, BH mass and charge are $M=1$, $Q=0.2$, respectively).}
\label{fig:ZnearStepPot}
\end{figure*}
\begin{figure*}[ht]
\begin{tabular}{ccc}
\includegraphics[width=0.45\textwidth,clip]{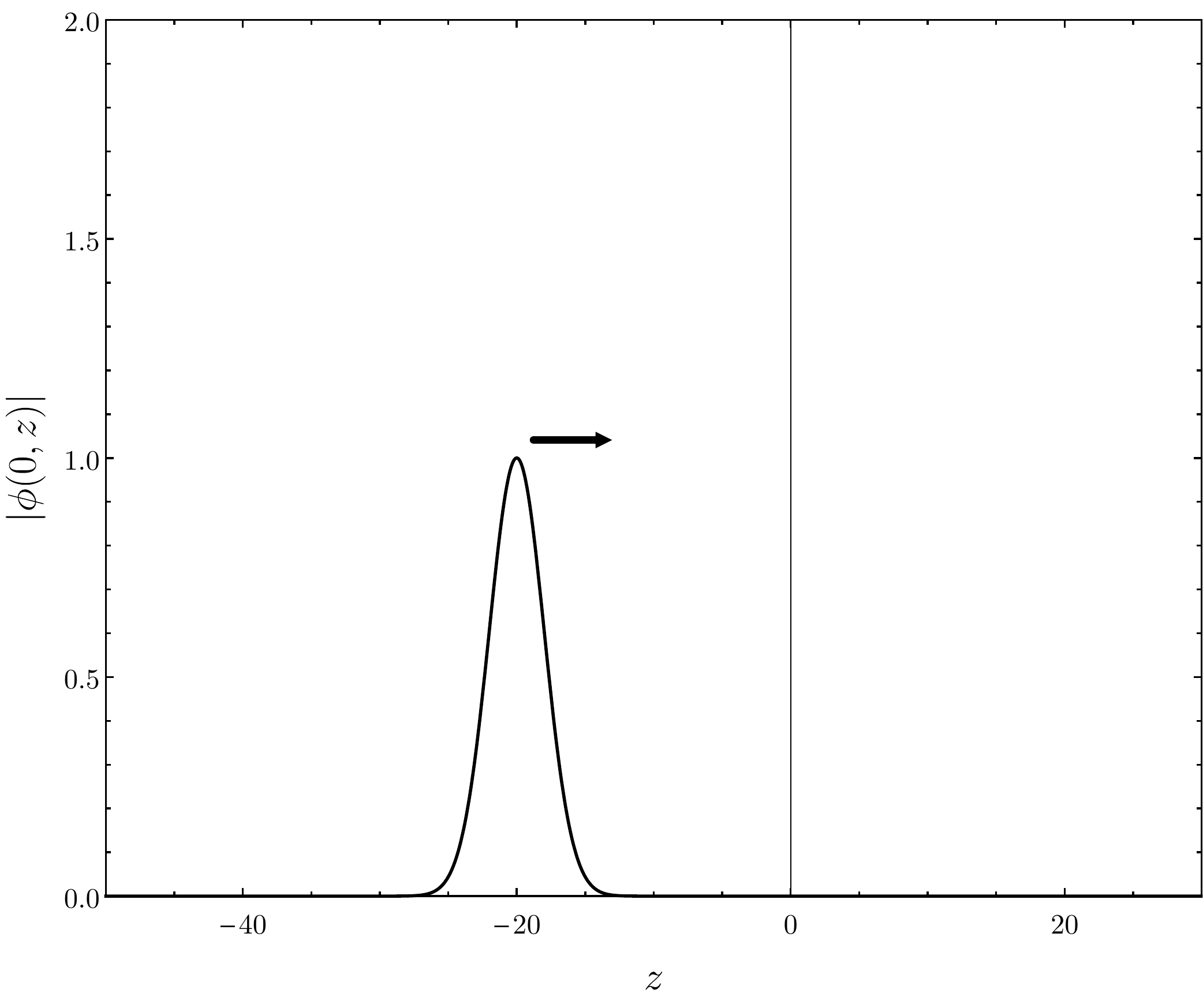}&
\includegraphics[width=0.45\textwidth,clip]{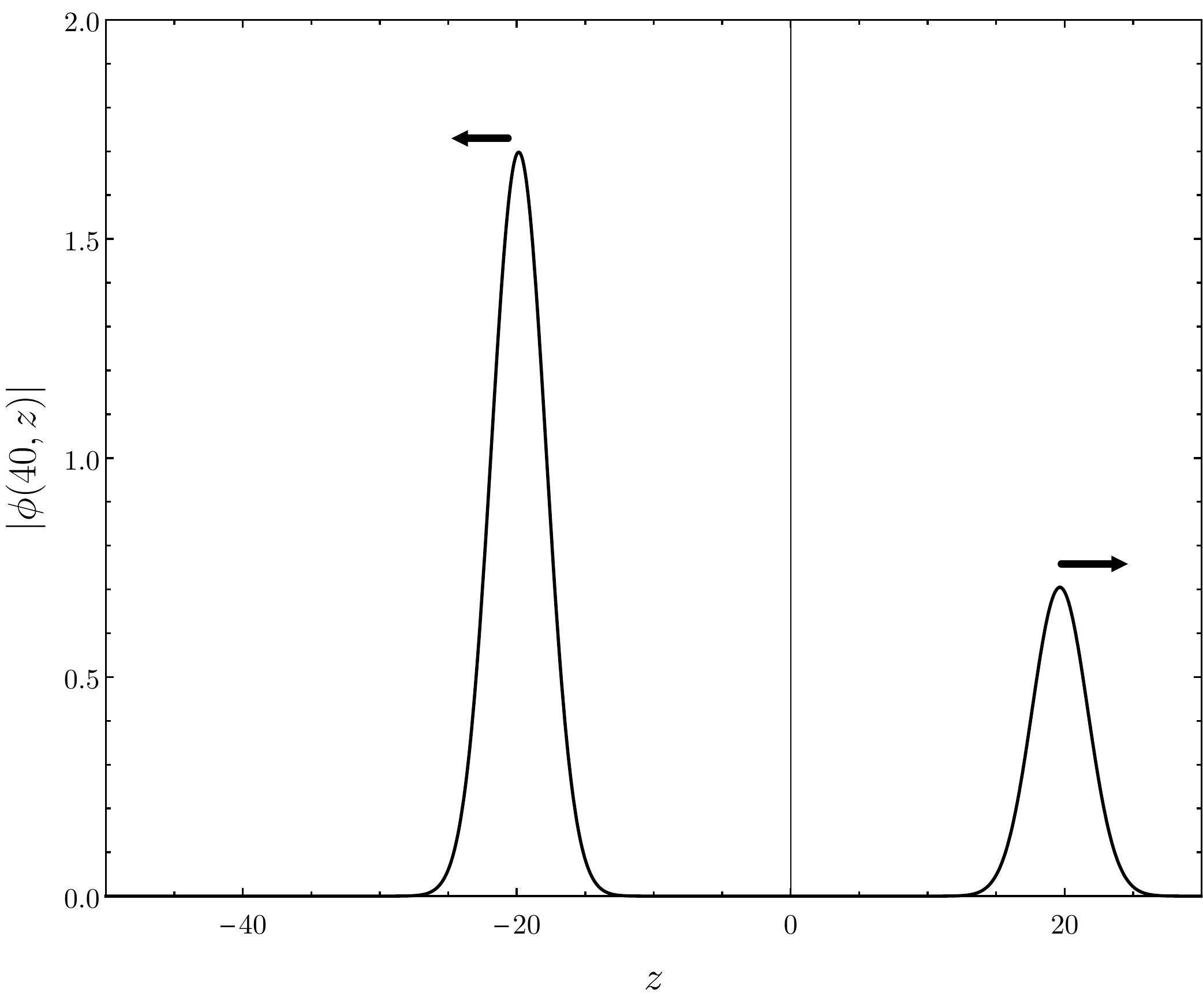}
\end{tabular}
	\caption{Wave packet solution at $t=0, 40$, for the left and right panel respectively. The initial data is given by Eqs.~\eqref{ic1}-\eqref{ic2} with $\omega_0=2$, $\sigma_z=2.24$ and $z_0=-20$. The potential has the form \eqref{nearstep} with $q\Phi_0=10$ and $L=0.001$. The figure on the left shows the form of the incident pulse before the scattering on the barrier. At $t=40$ the initial pulse scattered off the barrier, a fraction was reflected and a part penetrated inwards.}
	\label{fig:phi0num}
\end{figure*}
\begin{table}[H]
	\centering
	\caption{Amplification factors for the Klein and BH setup, for different width $\sigma_z$. For the Klein setup we used a coupling $q \Phi_0=10$, $L=0.001$ and $\omega_0=2$. Notice that, for this scattering problem, monochromatic waves of frequency $\omega=2$ have $Z\simeq2.778$. For the BH setup,  $M=1$, $Q=0.2$, $qQ=20$ and $\omega_0=2$. For this scattering problem, monochromatic waves of frequency $\omega=2$ have $Z\simeq1.955$.}
	\label{tab:ZScalarSigmanum}
	\begin{tabular}{ll||ll}
\\		\hline
\multicolumn{2}{c}{Klein} &\multicolumn{2}{c}{Black Hole}  \\ \hline
                          $\sigma_z$ & $Z$         & $\sigma_z$    &$Z$                    \\ \hline
		                        2.2      & 2.916       &  1.0         & 2.032 \\ 
		                        3.5      & 2.810       &  2.0         & 1.953 \\ 
		                        3.9      & 2.801       &  3.0         & 1.954 \\
		                        5.0      & 2.782       &  4.0         & 1.955\\ \hline
	\end{tabular}
\end{table} 
Our results are summarized in Figs.~\ref{fig:ZnearStepPot}~-~\ref{fig:phi0num} and in Table~\ref{tab:ZScalarSigmanum}.
Both the amplification factors and time evolution are consistent with those of the previous section.
In Table~\ref{tab:ZScalarSigmanum} we show some of the amplification factors obtained numerically for different values of $\sigma_z$, respectively for the near step barrier and RN BH. As expected, when $\sigma_z$ increases (the wave is more delocalized) the amplification factor of the pulse goes over to the value of the monochromatic waves.

\section{Superradiance and ergoregion instabilities in horizonless spacetimes}

\subsection{Horizons have nothing to do with superradiance}
%
\begin{figure*}[ht]
\begin{tabular}{cc}
\includegraphics[width=0.45\textwidth,clip]{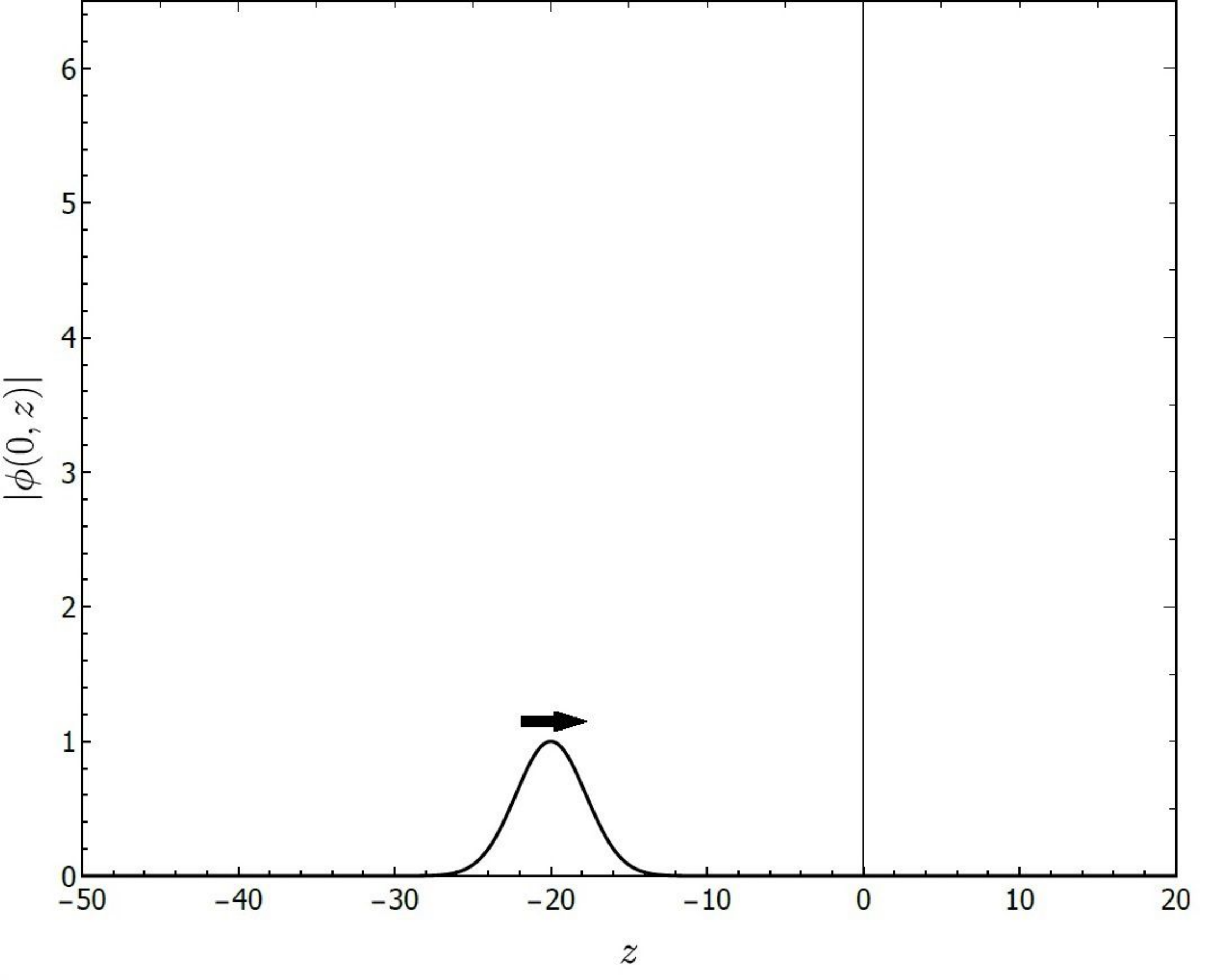}&
\includegraphics[width=0.45\textwidth,clip]{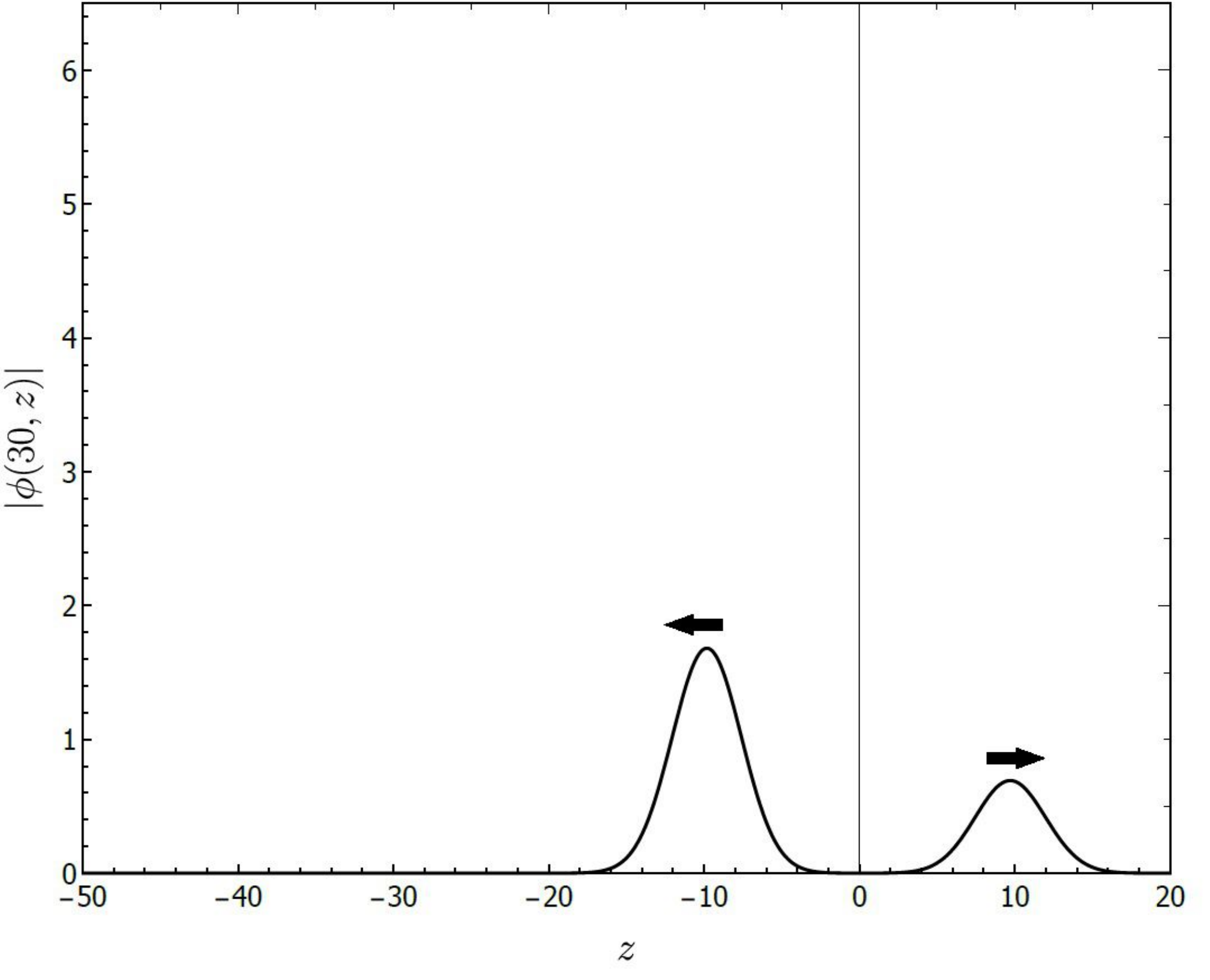}\\
\includegraphics[width=0.45\textwidth,clip]{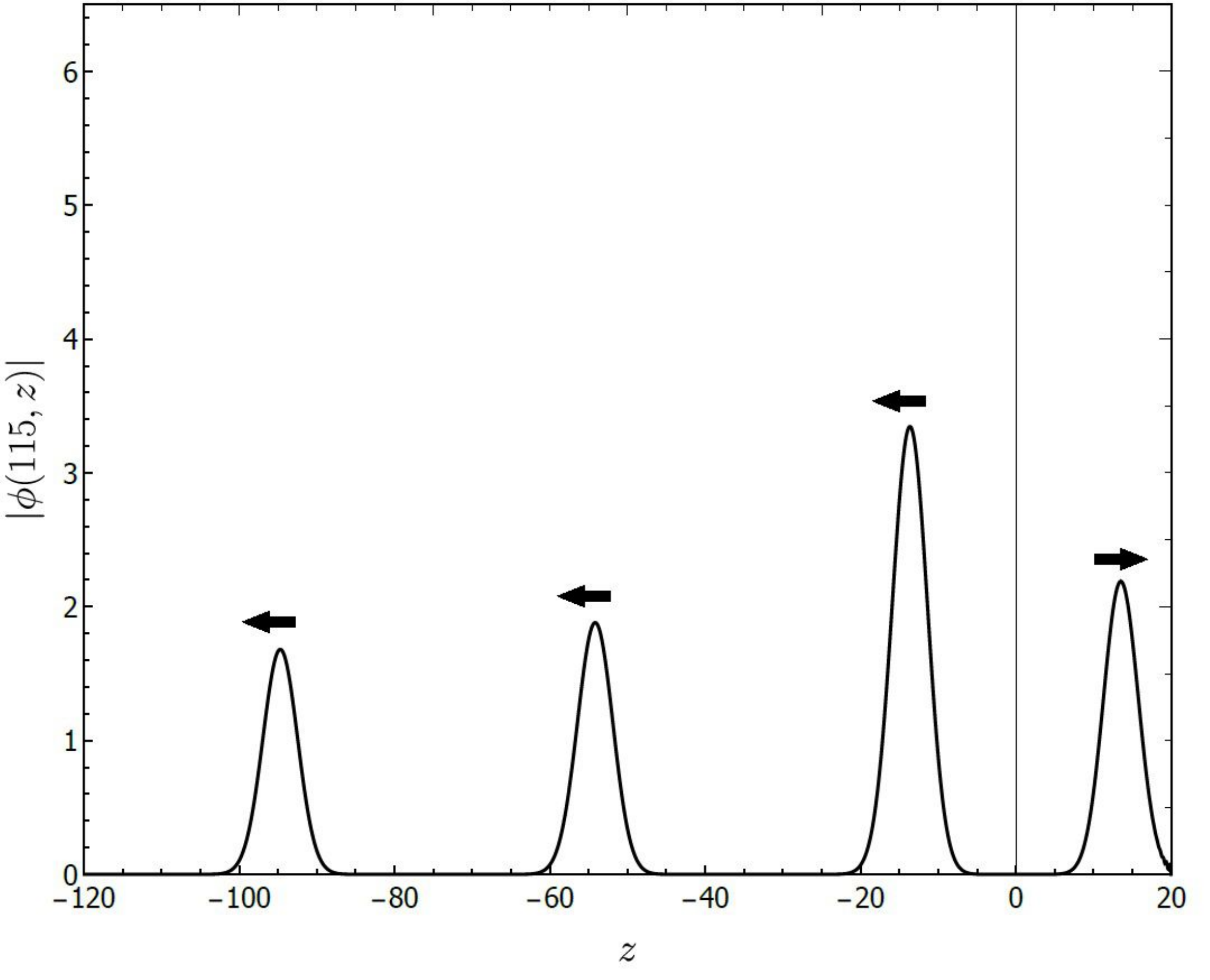}&
\includegraphics[width=0.44\textwidth,clip]{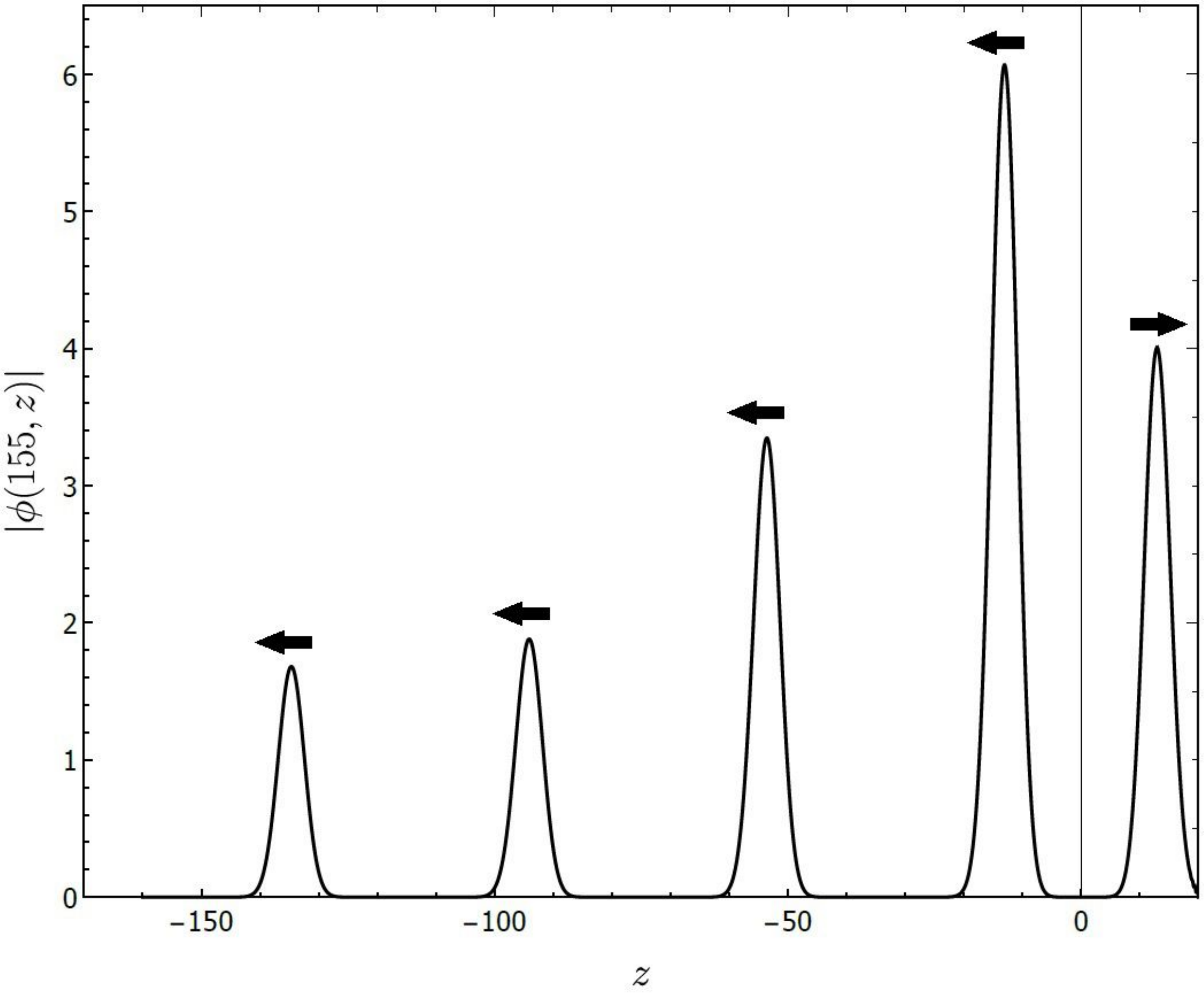}
\end{tabular}
\caption{Wave packet solution at $t=0, 30, 115, 155$, for the left and right panel respectively. The initial data is given by Eqs.~\eqref{ic1}-\eqref{ic2} with $\omega_0=2$, $\sigma_\omega=2.24$ and $z_0=-20$. The potential has the form \eqref{nearstep} with $q\Phi_0=10$ and $L=0.001$. There is now a mirror (implemented with a Dirichlet boundary condition) at $z=20$. This figure shows the form of the incident pulse before the scattering on the barrier. At $t=30$ the initial pulse scattered off the barrier, a fraction was reflected and a part penetrated inwards (just like in Fig.~\ref{fig:phi0num}). At latter times it is possible to see an analogue ergoregion instability (due to the successive superradiant amplification of the pulse confined in $0<z<20$). This instability is characterized by the emission of echoes of increasing amplitude.}
	\label{fig:phi0num1}
\end{figure*}
As we already discussed, boundary conditions are important ingredients in any evolution problem; however, when dealing with spacetimes with horizons,
the boundary condition at the horizon should be irrelevant: since waves take an infinite (coordinate) time to reach the horizon,
it has no causal contact with the region where ``dynamics is happening.'' Thus, boundary conditions should be irrelevant.
This seemingly harmless statement is at odds with recent results on the scattering of monochromatic waves, for which it is claimed that it is the boundary conditions on the fields that dictates superradiance~\cite{Richartz:2009mi}. It is thus generally believed (erroneously) that replacing an horizon by a hard surface would lead to no superradiance in BH geometries. In the following we show that pulses can be superradiantly amplified, independently of the boundary conditions imposed on the system.

We will focus on the simpler potential barrier Klein problem, but our results generalize to BH spacetimes in an obvious way.
Consider the Klein setup with the potential~\eqref{nearstep}, but this time with a mirror at $z_m\gg L$. This is implemented by imposing a Dirichlet boundary condition at $z_m$, \textit{i.e.} $\phi(t, z_m)=0$. We evolved Eq.~\eqref{scalarfe0} with initial conditions described by Eqs.~\eqref{ic1}-\eqref{ic2}, the results are shown in Fig.~\ref{fig:phi0num1}.

Our results show that for $z_m\gg \sigma_z$, the incident pulse is amplified {\it exactly in the same way} as when the mirror was absent (compare Figs.~\ref{fig:phi0num} and \ref{fig:phi0num1}).
Thus, wavepackets are amplified regardless of the mirror or of the existence of an horizon in the spacetime.
This can be understood by locality arguments, since when the pulse is scattered it has not yet ``felt'' the mirror.

In the absence of a mirror (or, in BH spacetimes, in the presence of horizons) the initial pulse scatters off the barrier and gives rise to (i) a (superradiantly) amplified signal and (ii)
a transmitted signal which travels down the barrier (into the horizon). When surfaces are present, however, the transmitted pulse is eventually forced to return where it leads to a 
``echoe'' pulse and another transmitted signal. Because the signal within the barrier has ``negative-energy'', each interaction leads to a growth of this pulse and therefore
to ``echoes'' increasing in amplitude~\cite{Cardoso:2016rao,Cardoso:2017njb,Cardoso:2017cqb}. Thus, the absence of horizons is also deeply connected to superradiant instabilities. 
The increase in amplitude of successive echoes is also translated into the existence of unstable quasi-normal modes (QNMs) of the system. We have computed the QNMs of these systems and found good agreement with the results from time evolutions.

As we stressed, the results for BH spacetimes are qualitatively the same once a reflecting surface is placed outside but close to the horizon. This implies that the necessary condition for superradiant amplification of localized pulses is the same as for energy extraction by means of the Penrose process -- the existence of an ergoregion (or, more generally, a region of negative energy). In other words, horizons are not necessary. Furthermore, this kind of setups also leads to superradiant instabilities (with the production of echoes of increasing amplitude in time). 
There is in fact a generic proof of this fact: asymptotically flat, horizonless geometries with ergoregions are unstable~\cite{Cardoso:2017njb,Cardoso:2017cqb}. This result is in complete agreement with our findings and we see that these instabilities are due to superradiant amplification of fields confined in the ergoregions.

How, then, are our results compatible with previous investigations~\cite{Richartz:2009mi} in the frequency domain? Imagine that we apply the procedure of Section~\ref{sec:fourier} to the same step potential, but with a reflecting mirror at some $\widetilde{z}\gg0$ (\textit{e.g.}, imposing Dirichlet or Neumann boundary conditions at $\widetilde{z}$). In this case, it is easy to see that $|R|=|I|$. Furthermore, since Eq.~\eqref{Z} also holds for this setup, it implies $Z=1$. One can misinterpret this result, by concluding that no superradiant amplification occurs in these setups, and that the number of reflected particles at infinity is exactly the same as the number of particles that was sent initially. It turns out that this is not the correct interpretation of the result. What it indicates is that only stable solutions $\phi(t,z)$ can be written in the form \eqref{phiwp1}, these having $Z=1$. As we have just seen in the time evolutions, instabilities {\it are} triggered in the absence of horizons, leading to non-normalizable solutions, which cannot possibly fit the framework of those studies~\cite{Richartz:2009mi}. Nevertheless, such an approach is still appropriate to analyze the scattering of pulses containing non-superradiant modes (which do not trigger instabilities), and the results coincide with the ones obtained through time evolutions. 

 \subsection{Instabilities with point particles}
We will now show that ergoregion instabilities can occur also with classical particles, once Penrose-like processes are allowed to occur in a horizonless spacetime.
Let us consider a rotating (Kerr) BH. This solution is known to have an ergoregion where point particles can have negative energy with respect to observers at infinity~\cite{Penrose:1971uk,Brito:2015oca,gravitation}. This fact makes Kerr BHs prone to energy extraction by classical particles through the so-called Penrose process (see section~\ref{subsec:penprocess}).
 
 Now imagine the following energy extraction process happening in the equatorial plane of an extremal Kerr BH~\footnote{For simplicity we specialize to an extremal BH. Our results should have a straightforward generalization.}:
 
 \textbf{1.}\quad A particle (denoted by $0$) is left from infinity with $(E_0/\mu_0)\equiv \epsilon_0=1$ and $(L_0/\mu_0)\equiv l_0$, where $E_0$ and $L_0$ are, respectively, the energy and angular momentum of the particle (with respect to an observer at infinity) and $\mu_0$ its rest mass.
 
 \textbf{2.}\quad This initial particle stops moving radially at $\widetilde{r}$ ($\dot{r_0}(\widetilde{r})=0$), where it decays in two other particles ($1$ and $2$) with equal rest mass $\mu_1=\mu_2=x \mu_0$. It is easy to show that (\textit{e.g}, Eq. $4.27$ of Ref.~\cite{Brito:2015oca})
 \begin{equation}
 	l_0=\frac{2 M^2-\left(\widetilde{r}-M\right)\sqrt{2 \widetilde{r} M}}{2 M-\widetilde{r}}\,.
 \end{equation}
If one of these product particles is to reach infinity, the decay must occur outside the BH horizon ($\widetilde{r}>r_+=M$). This condition implies $$\frac{E_0}{L_0}=\frac{1}{l_0}<\frac{1}{2 r_+}=\Omega_+\,,$$ which is analogous to the superradiant condition $(\omega/m)<\Omega_+$. On the other hand, if energy extraction is to happen, the particle $0$ must decay inside the ergoregion ($\widetilde{r}<2M$, since, in the equatorial plane of an extremal Kerr BH, the ergoregion is a circle of radius $r=2 M$). The latter condition implies $$\frac{E_0}{L_0}=\frac{1}{l_0}>\frac{2}{5 M}\,.$$ Then, the superradiant condition is necessary, but not sufficient for this energy extraction process to happen. The inequality $(0.4/M)<(E_0/L_0)<\Omega_+$ must hold.
 
 Assuming $\dot{r_1}(\widetilde{r})=\dot{r_2}(\widetilde{r})=0$ right after the decay, one obtain (\textit{e.g}, Eq. $4.28$ of Ref.~\cite{Brito:2015oca})
 \begin{align}\label{li}
 	l_i=\frac{2 \epsilon_i M^2-\left(\widetilde{r}-M\right)\sqrt{\widetilde{r}\left(2 M+\left(\epsilon_i^2-1\right) \widetilde{r}\right)}}{2 M-\widetilde{r}}\, ,
 \end{align}
 with $i=1,2$.
 By conservation laws
 \begin{equation}
 	\epsilon_1+\epsilon_2=\frac{1}{x} \epsilon_0=\frac{1}{x} \,,\quad l_1+l_2=\frac{1}{x} l_0 \,,
 \end{equation} 
implying
\begin{align}
	\epsilon_1&=\frac{1}{2 x}\left(1-\sqrt{\frac{2 M}{\widetilde{r}}(1-4 x^2)}\right)\,, \label{epsi1} \\
    \epsilon_2&=\frac{1}{2 x}\left(1+\sqrt{\frac{2 M}{\widetilde{r}}(1-4 x^2)}\right)\,, 
\end{align}
where we use even subscripts for positive energy particles and odd subscripts for the negative ones.  

Until now we have described only the standard Penrose process. Furthermore, we recover the well-known upper bound for the efficiency of the (Penrose) energy extraction~\cite{Brito:2015oca}
\begin{equation}
	\text{max}\left(\frac{E_2}{E_0}\right)=\text{max}\left(\epsilon_2 x\right)=\frac{1+\sqrt{2}}{2}\, ,
\end{equation}
obtained for $\widetilde{r}=M$ and $x=0$. 

Let us consider now the same process in the neighborhood of a very compact (rotating) star of radius 
\be
r_{\text{star}}=r_+(1+\delta)\,,
\ee
with $\delta\ll 1$. We assume that its exterior is described by an extremal Kerr metric and that the collisions of the point particles with the compact object are elastic, conserving their energy and angular momentum (we are assuming the mass of the particles to be negligible compared with the mass of the star). Consider that particle 1 falls into the star (we will return to this assumption later). 
It is well-known that the equations of motion of point particles (in general relativity) are time-reversible. 
Thus, after colliding with the surface of the star, particle 1 moves away from it symmetrically to the way it had approached the surface before, and it stops moving radially at the same $\widetilde{r}$. In other words, the path described by particle 1 \textit{before collision} (approaching $r_{\text{star}}$ from $\widetilde{r}$) is symmetric to the path \textit{after collision} (moving away from $r_{\text{star}}$ to $\widetilde{r}$).

Now we imagine that the process continues with:

\textbf{3.}\quad The negative energy particle $1$ returns to $\widetilde{r}$ \textit{after the collision}, where it stops moving radially ($\dot{r_1}(\widetilde{r})=0$) and it decays in two other particles ($3$ and $4$) with equal rest mass $\mu_3=\mu_4=x \mu_1= x^2 \mu_0$. 

\begin{figure}[h!]
	\centering
	\includegraphics[width=0.8\linewidth]{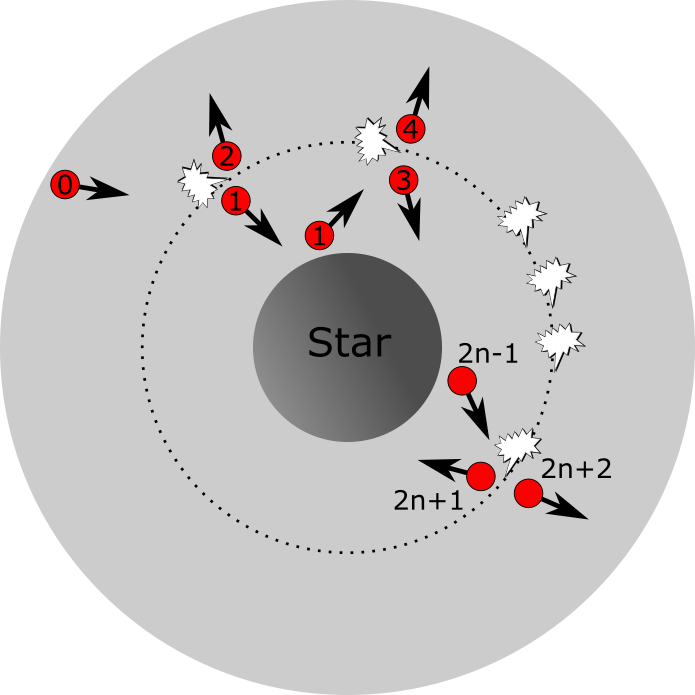}
	\caption{Penrose process cascade in the neighborhood of a very compact (rotating) star. A particle decays in two inside the ergoregion, at a radius $\widetilde{r}$ (the dashed circumference). 
	The ingoing particle is made to reflect off the surface of the star and decay again at the same radius. The process continues indefinitely and gives rise to an ergoregion (exponential) instability. }
	\label{fig:Pencasc}
\end{figure}

Assuming $\dot{r_3}(\widetilde{r})=\dot{r_4}(\widetilde{r})=0$ right after the decay, one can show that Eq.~\eqref{li} also holds for particles $3$ and $4$. By conservation laws
\begin{align}
	\epsilon_3 +\epsilon_4=\frac{1}{x}\epsilon_1 \, , \quad l_3+l_4=\frac{1}{x}l_1\, ,
\end{align}
allowing us to find (at least numerically) $\epsilon_3$ and $\epsilon_4$ as functions of $\epsilon_1$ (which is known from Eq.~\eqref{epsi1}). Moreover, one can show that one of the product particles has negative energy and the other has positive energy. Following our subscript convention, particle $3$ has negative energy (falling into the star), whereas particle $4$ has positive energy. Then, particle $3$ collides with the star and the process happens over and over again (see Fig.~\ref{fig:Pencasc}). So, for $n \in \mathbb{N}$:

\textbf{4.}\quad The negative energy particle $2 n -1$ returns to $\widetilde{r}$ \textit{after the collision}, where it stops moving radially ($\dot{r}_{2n-1}(\widetilde{r})=0$) and it decays in two other particles ($2 n+1$ and $2 n+2$) with equal rest mass $\mu_{2 n+1}=\mu_{2 n+2}=x \mu_{2 n -1}= x^{n+1} \mu_0$. 

Assuming $\dot{r}_{2n+1}(\widetilde{r})=\dot{r}_{2n+2}(\widetilde{r})=0$ right after the decay, one can show that Eq.~\eqref{li} also holds for particles $2n+1$ and $2n+2$. By conservation laws
\begin{align}
\epsilon_{2n+1} +\epsilon_{2n+2}=\frac{1}{x}\epsilon_{2n-1} \, , \quad l_{2n+1}+l_{2n+2}=\frac{1}{x}l_{2n-1}\, ,
\end{align}
allowing us to find (at least numerically) $\epsilon_{2n+1}$ and $\epsilon_{2n+2}$ as functions of $\epsilon_{2n-1}$ (which is known cyclically from Eq.~\eqref{epsi1}). Solving the equations of motion numerically it is possible to see that the time (with respect to a stationary observer at infinity) that a particle takes to fall into the star is $\Delta t \sim M/\delta$. Thus, this time scale is not very sensible to neither $\widetilde{r}$ nor $x$.

\begin{figure}[h!]
\centering
\includegraphics[width=1\linewidth]{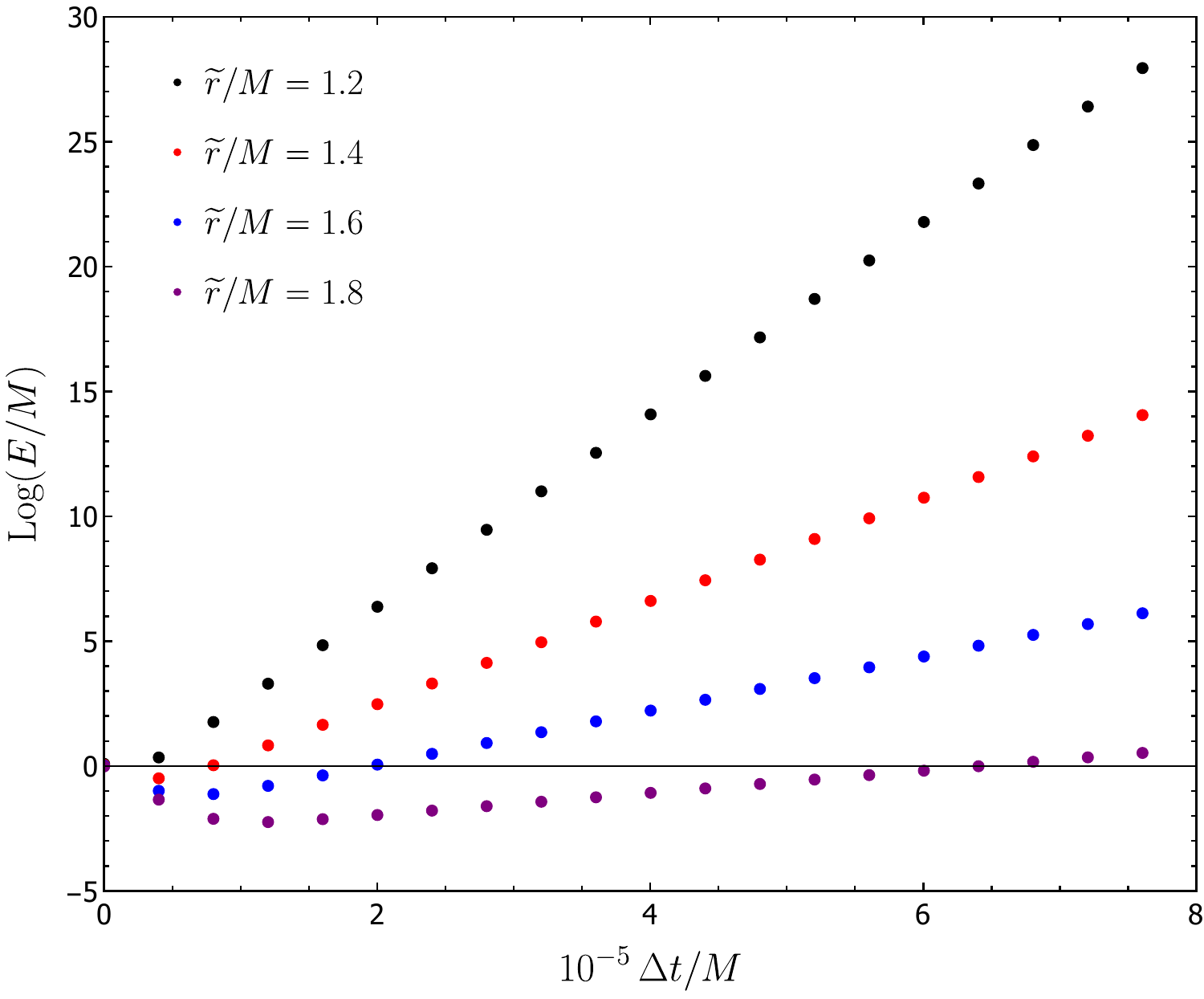}
\caption{Exponential growth of the energy of first 20 emitted (positive energy) particles for different values of $\widetilde{r}/M$, with $x=0.2$ and $\delta=10^{-4}$.}
\label{fig:Pensrr}
\end{figure}

\begin{figure}[h!]
	\centering
	\includegraphics[width=1\linewidth]{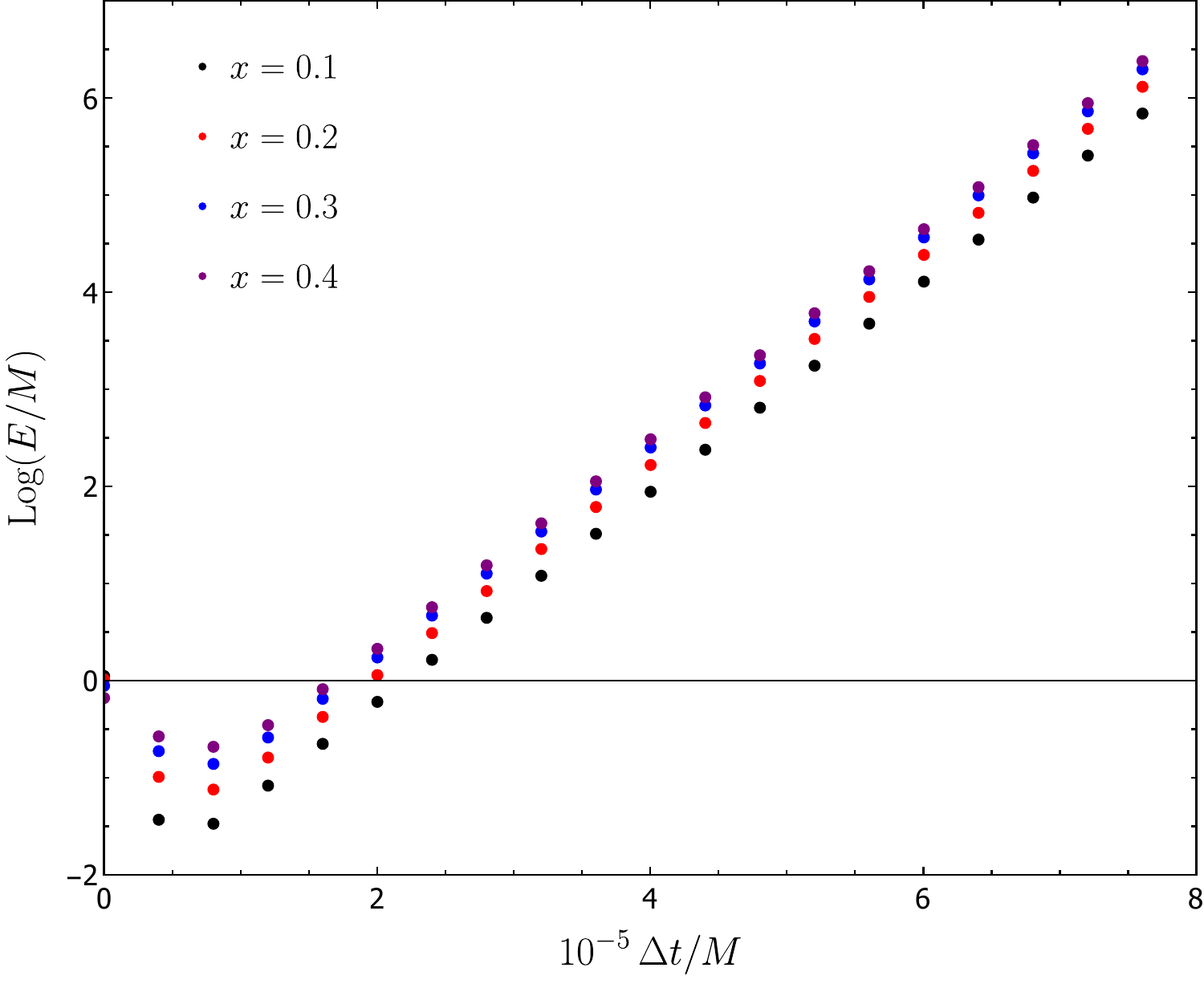}
	\caption{Exponential growth of the energy of first 20 emitted (positive energy) particles for different values of $x$, with $\widetilde{r}/M=1.6$ and $\delta=10^{-4}$. }
	\label{fig:Pensrx}
\end{figure}

In Figs.~\ref{fig:Pensrr}--\ref{fig:Pensrx} we plot the energy of the first 20 (positive energy) product particles just after the decay of their parent particle at $\widetilde{r}$. It is clear from the figures that the energy of the emitted particles grows exponentially in time. Thus, this Penrose process cascade gives rise to an instability, characterized by the emission of particles of growing energy to infinity. This instability can be thought as particle analogue of a superradiant instability (which is accompanied by the emission of echoes of growing amplitude to infinity).

At a certain point in describing the process we assumed that every negative-energy particle falls into the star after being created. In fact, this is not always the case. Some of these particles can remain in orbit around the star, the majority being circular (of radius $\widetilde{r}$) and unstable (the ISCO is $r=9 M$ for retrograde orbits in extremal Kerr). Since these orbits are unstable, we assume that eventually the particles will fall into the star (in the same time scale $\Delta t \sim M/\delta$). In the same way, not all the emitted positive energy particles reach infinity: some of them fall into the star; others stay in closed orbits around it, the majority being circular (of radius $\widetilde{r}$) and stable (the ISCO is $r=M$ for prograde orbits in extremal Kerr). Doing a numerical analysis of the first 50 (positive energy) emitted particles, we observe that the higher the $\widetilde{r}/M$, the larger the number of particles reaching infinity. On the other hand, the number of particles reaching infinity does not depend strongly on $x$. Thus, this process can power the formation of a disc around a compact (rotating) star. In fact, it is not hard to find numerically the (particle) density profile of this disc.

We expect the main conclusions of this section to hold under more general conditions: whenever there is a negative energy region and horizons are not present. Under these conditions, a (generalized) Penrose process cascade may be possible, giving rise to an instability and powering the formation of some structure outside the surface.

\section{Conclusions}

In this work we explored several connections between two energy extraction processes: the Penrose process and superradiance. We showed that, although the two processes are different in nature, they can happen in the same situations (whenever a region of negative energy is present), and both can give rise to instabilities (if, additionaly, horizons are not present).

To conclude, we briefly summarize our results and point out possible directions of future research.

\textbf{1.}\quad We reviewed the scattering of scalar and Dirac fields on electrostatic potential barriers and on RN BHs: showing that scalar fields are prone to superradiant amplification, whereas Dirac fields are not~\cite{Greiner1985,MANOGUE1988261,Brito:2015oca,Lee:1977gk}. In particular, we used the spherical symmetry of RN geometry to separate the Dirac equation, proving then that these fields cannot be amplified on RN. As far as we are aware, this result was known only as a limit of the more general Kerr-Newman geometry~\cite{Lee:1977gk}.

\textbf{2.}\quad We gave a physical motivation for the existence of bosonic fermion condensates (\textit{i.e.} systems of fermions with bosonic behavior as a whole), which could be (superradiantly) amplified. Furthermore, we provided a non-linear Dirac theory as a proof of principle; this theory admits superradiant scattering of monochromatic solutions on a Klein setup. 
Open issues concern direct time-evolution of wavepackets and an extension to curved spacetimes. The problem of finding a Dirac theory on curved spacetime that admits amplification does not seem to be straightforward. To solve some of these problems, we believe it is important to understand which boundary conditions should be used; because of the non-linearity, it is not obvious that the concept of group velocity is well defined here. Nevertheless, in our treatment, we assumed it to be.  

\textbf{3.}\quad We employed two different methods to analyze the scattering of pulses on Klein setup and RN geometry: a Fourier-domain calculation and a fully numerical evolution in time. We proved that the results obtained for the scattering of pulses are compatible with the monochromatic ones (in the quasi-monochromatic limit).
Furthermore, we showed that boundary conditions do not dictate superradiance. In particular, superradiance is possible whenever the Penrose process is: when a region of negative energy is present. 

\textbf{4.}\quad We showed numerically that when there is a region of negative energy and horizons are not present there exists a superradiant instability, characterized by the emission of echoes of growing amplitude to infinity~\cite{Cardoso:2016rao,Cardoso:2017njb,Cardoso:2017cqb}.

\textbf{5.}\quad We proved that if a compact (rotating) star has an ergoregion, a Penrose process cascade can give rise to an instability, characterized by the emission of particles of growing energy to infinity and powering the formation of a disc around the star. This instability can be thought of as particle analogue of the superradiant one, and, in particular, it can be shown to grow on 
similar timescales. The end state, and possible structure formation outside the star needs further investigation.

\begin{acknowledgments}
RV was supported by a graduate research fellowship
from the FCT/Portugal project UID/FIS/00099/2013 and by the FCT PhD scholarship SFRH/BD/128834/2017.
V. C. is indebted to Kinki University in Osaka for hospitality while the last stages of this work were being completed.
The authors acknowledge financial support provided under the European Union's H2020 ERC Consolidator Grant ``Matter and strong-field gravity: New frontiers in Einstein's theory'' grant agreement no. MaGRaTh--646597. Research at Perimeter Institute is supported by the Government of Canada through Industry Canada and by the Province of Ontario through the Ministry of Economic Development $\&$
Innovation.
This work has received funding from the European Union's Horizon 2020 research and innovation programme under the 
Marie Sk\l odowska-Curie grant agreement No 690904.
This article is based upon work from COST Action CA16104 ``GWverse'', supported by COST (European Cooperation in Science and Technology).
\end{acknowledgments}

\bibliographystyle{apsrev4}

\begin{thebibliography}{62}%
\makeatletter
\providecommand \@ifxundefined [1]{%
 \@ifx{#1\undefined}
}%
\providecommand \@ifnum [1]{%
 \ifnum #1\expandafter \@firstoftwo
 \else \expandafter \@secondoftwo
 \fi
}%
\providecommand \@ifx [1]{%
 \ifx #1\expandafter \@firstoftwo
 \else \expandafter \@secondoftwo
 \fi
}%
\providecommand \natexlab [1]{#1}%
\providecommand \enquote  [1]{``#1''}%
\providecommand \bibnamefont  [1]{#1}%
\providecommand \bibfnamefont [1]{#1}%
\providecommand \citenamefont [1]{#1}%
\providecommand \href@noop [0]{\@secondoftwo}%
\providecommand \href [0]{\begingroup \@sanitize@url \@href}%
\providecommand \@href[1]{\@@startlink{#1}\@@href}%
\providecommand \@@href[1]{\endgroup#1\@@endlink}%
\providecommand \@sanitize@url [0]{\catcode `\\12\catcode `\$12\catcode
  `\&12\catcode `\#12\catcode `\^12\catcode `\_12\catcode `\%12\relax}%
\providecommand \@@startlink[1]{}%
\providecommand \@@endlink[0]{}%
\providecommand \url  [0]{\begingroup\@sanitize@url \@url }%
\providecommand \@url [1]{\endgroup\@href {#1}{\urlprefix }}%
\providecommand \urlprefix  [0]{URL }%
\providecommand \Eprint [0]{\href }%
\providecommand \doibase [0]{http://dx.doi.org/}%
\providecommand \selectlanguage [0]{\@gobble}%
\providecommand \bibinfo  [0]{\@secondoftwo}%
\providecommand \bibfield  [0]{\@secondoftwo}%
\providecommand \translation [1]{[#1]}%
\providecommand \BibitemOpen [0]{}%
\providecommand \bibitemStop [0]{}%
\providecommand \bibitemNoStop [0]{.\EOS\space}%
\providecommand \EOS [0]{\spacefactor3000\relax}%
\providecommand \BibitemShut  [1]{\csname bibitem#1\endcsname}%
\let\auto@bib@innerbib\@empty
\bibitem [{\citenamefont {Cardoso} \emph
  {et~al.}(2016{\natexlab{a}})\citenamefont {Cardoso}, \citenamefont {Coutant},
  \citenamefont {Richartz}, and \citenamefont {Weinfurtner}}]{Cardoso:2016zvz}%
  \BibitemOpen
  \bibfield  {author} {\bibinfo {author} {\bibfnamefont {V.}~\bibnamefont
  {Cardoso}}, \bibinfo {author} {\bibfnamefont {A.}~\bibnamefont {Coutant}},
  \bibinfo {author} {\bibfnamefont {M.}~\bibnamefont {Richartz}},  and \bibinfo
  {author} {\bibfnamefont {S.}~\bibnamefont {Weinfurtner}}, }\href {\doibase
  10.1103/PhysRevLett.117.271101} {\bibfield  {journal} {\bibinfo  {journal}
  {\emph {Phys. Rev. Lett.}} }\textbf {\bibinfo {volume} {117}}, \bibinfo
  {pages} {271101} (\bibinfo {year} {2016}{\natexlab{a}})}, \Eprint
  {http://arxiv.org/abs/1607.01378} {arXiv:1607.01378}\BibitemShut {NoStop}%
\bibitem [{\citenamefont {Dicke}(1954)}]{Dicke}%
  \BibitemOpen
  \bibfield  {author} {\bibinfo {author} {\bibfnamefont {R.H.} \bibnamefont
  {Dicke}}, }\href {\doibase 10.1103/PhysRev.93.99} {\bibfield  {journal}
  {\bibinfo  {journal} {\emph {Phys. Rev.}} }\textbf {\bibinfo {volume} {93}},
  \bibinfo {pages} {99} (\bibinfo {year} {1954})}\BibitemShut {NoStop}%
\bibitem [{\citenamefont {Gross} and \citenamefont
  {Haroche}(1982)}]{GROSS1982301}%
  \BibitemOpen
  \bibfield  {author} {\bibinfo {author} {\bibfnamefont {M.}~\bibnamefont
  {Gross}} and \bibinfo {author} {\bibfnamefont {S.}~\bibnamefont {Haroche}},
  }\href {\doibase http://dx.doi.org/10.1016/0370-1573(82)90102-8} {\bibfield
  {journal} {\bibinfo  {journal} {\emph {Physics Reports}} }\textbf {\bibinfo
  {volume} {93}}, \bibinfo {pages} {301 } (\bibinfo {year} {1982})}\BibitemShut
  {NoStop}%
\bibitem [{\citenamefont {Manogue}(1988)}]{MANOGUE1988261}%
  \BibitemOpen
  \bibfield  {author} {\bibinfo {author} {\bibfnamefont {C.A.} \bibnamefont
  {Manogue}}, }\href {\doibase http://dx.doi.org/10.1016/0003-4916(88)90167-4}
  {\bibfield  {journal} {\bibinfo  {journal} {\emph {Annals of Physics}}
  }\textbf {\bibinfo {volume} {181}}, \bibinfo {pages} {261 } (\bibinfo {year}
  {1988})}\BibitemShut {NoStop}%
\bibitem [{\citenamefont {Greiner}(1985)}]{Greiner1985}%
  \BibitemOpen
  \bibfield  {author} {\bibinfo {author} {\bibfnamefont {W.}~\bibnamefont
  {Greiner}}, }\enquote {\bibinfo {title} {Quantum electrodynamics of strong
  fields},} in \href {\doibase 10.1007/3-540-15653-4_3} {\emph {\bibinfo
  {booktitle} {Hadrons and Heavy Ions: Proceedings of the Summer School Held at
  the University of Cape Town January 16--27, 1984}}}, \bibinfo {editor}
  {edited by \bibinfo {editor} {\bibfnamefont {W.D.} \bibnamefont {Heiss}}}
  (\bibinfo  {publisher} {Springer Berlin Heidelberg}, \bibinfo {address}
  {Berlin, Heidelberg}, \bibinfo {year} {1985})\BibitemShut {NoStop}%
\bibitem [{\citenamefont {Bekenstein} and \citenamefont
  {Schiffer}(1998)}]{Bekenstein:1998nt}%
  \BibitemOpen
  \bibfield  {author} {\bibinfo {author} {\bibfnamefont {J.D.} \bibnamefont
  {Bekenstein}} and \bibinfo {author} {\bibfnamefont {M.}~\bibnamefont
  {Schiffer}}, }\href {\doibase 10.1103/PhysRevD.58.064014} {\bibfield
  {journal} {\bibinfo  {journal} {\emph {Phys. Rev.}} }\textbf {\bibinfo
  {volume} {D58}}, \bibinfo {pages} {064014} (\bibinfo {year} {1998})}, \Eprint
  {http://arxiv.org/abs/gr-qc/9803033} {arXiv:gr-qc/9803033}\BibitemShut
  {NoStop}%
\bibitem [{\citenamefont {Zeldovich}(1971)}]{Yakov}%
  \BibitemOpen
  \bibfield  {author} {\bibinfo {author} {\bibfnamefont {Y.B.} \bibnamefont
  {Zeldovich}}, }\href@noop {} {\bibfield  {journal} {\bibinfo  {journal}
  {\emph {Prisma Zh. Eksp. Teor. Fiz}} }\textbf {\bibinfo {volume}
  {\textbf{14}}}, \bibinfo {pages} {270} (\bibinfo {year} {1971})}\BibitemShut
  {NoStop}%
\bibitem [{\citenamefont {Brito} \emph
  {et~al.}(2015{\natexlab{a}})\citenamefont {Brito}, \citenamefont {Cardoso},
  and \citenamefont {Pani}}]{Brito:2015oca}%
  \BibitemOpen
  \bibfield  {author} {\bibinfo {author} {\bibfnamefont {R.}~\bibnamefont
  {Brito}}, \bibinfo {author} {\bibfnamefont {V.}~\bibnamefont {Cardoso}},  and
  \bibinfo {author} {\bibfnamefont {P.}~\bibnamefont {Pani}}, }\href {\doibase
  10.1007/978-3-319-19000-6} {\bibfield  {journal} {\bibinfo  {journal} {\emph
  {Lect. Notes Phys.}} }\textbf {\bibinfo {volume} {906}}, \bibinfo {pages}
  {pp.1} (\bibinfo {year} {2015}{\natexlab{a}})}, \Eprint
  {http://arxiv.org/abs/1501.06570} {arXiv:1501.06570}\BibitemShut {NoStop}%
\bibitem [{\citenamefont {Kim}(2008)}]{Kim}%
  \BibitemOpen
  \bibfield  {author} {\bibinfo {author} {\bibfnamefont {H.}~\bibnamefont
  {Kim}}, }\href {http://stacks.iop.org/1475-7516/2008/i=11/a=007} {\bibfield
  {journal} {\bibinfo  {journal} {\emph {Journal of Cosmology and Astroparticle
  Physics}} }\textbf {\bibinfo {volume} {2008}}, \bibinfo {pages} {007}
  (\bibinfo {year} {2008})}\BibitemShut {NoStop}%
\bibitem [{\citenamefont {Misner}(1972)}]{Misner:1972kx}%
  \BibitemOpen
  \bibfield  {author} {\bibinfo {author} {\bibfnamefont {C.W.} \bibnamefont
  {Misner}}, }\href {\doibase 10.1103/PhysRevLett.28.994} {\bibfield  {journal}
  {\bibinfo  {journal} {\emph {Phys. Rev. Lett.}} }\textbf {\bibinfo {volume}
  {28}}, \bibinfo {pages} {994} (\bibinfo {year} {1972})}\BibitemShut {NoStop}%
\bibitem [{\citenamefont {Teukolsky}(1972)}]{Teukolsky:1972my}%
  \BibitemOpen
  \bibfield  {author} {\bibinfo {author} {\bibfnamefont {S.A.} \bibnamefont
  {Teukolsky}}, }\href {\doibase 10.1103/PhysRevLett.29.1114} {\bibfield
  {journal} {\bibinfo  {journal} {\emph {Phys. Rev. Lett.}} }\textbf {\bibinfo
  {volume} {29}}, \bibinfo {pages} {1114} (\bibinfo {year} {1972})}\BibitemShut
  {NoStop}%
\bibitem [{\citenamefont {Teukolsky} and \citenamefont
  {Press}(1974)}]{Teukolsky:1974yv}%
  \BibitemOpen
  \bibfield  {author} {\bibinfo {author} {\bibfnamefont {S.A.} \bibnamefont
  {Teukolsky}} and \bibinfo {author} {\bibfnamefont {W.H.} \bibnamefont
  {Press}}, }\href {\doibase 10.1086/153180} {\bibfield  {journal} {\bibinfo
  {journal} {\emph {Astrophys. J.}} }\textbf {\bibinfo {volume} {193}},
  \bibinfo {pages} {443} (\bibinfo {year} {1974})}\BibitemShut {NoStop}%
\bibitem [{\citenamefont {Arvanitaki} \emph {et~al.}(2016)\citenamefont
  {Arvanitaki}, \citenamefont {Baryakhtar}, \citenamefont {Dimopoulos},
  \citenamefont {Dubovsky}, and \citenamefont {Lasenby}}]{Arvanitaki:2016qwi}%
  \BibitemOpen
  \bibfield  {author} {\bibinfo {author} {\bibfnamefont {A.}~\bibnamefont
  {Arvanitaki}}, \bibinfo {author} {\bibfnamefont {M.}~\bibnamefont
  {Baryakhtar}}, \bibinfo {author} {\bibfnamefont {S.}~\bibnamefont
  {Dimopoulos}}, \bibinfo {author} {\bibfnamefont {S.}~\bibnamefont
  {Dubovsky}},  and \bibinfo {author} {\bibfnamefont {R.}~\bibnamefont
  {Lasenby}}, }\href@noop {} {  (\bibinfo {year} {2016})}, \Eprint
  {http://arxiv.org/abs/1604.03958} {arXiv:1604.03958}\BibitemShut {NoStop}%
\bibitem [{\citenamefont {Brito} \emph
  {et~al.}(2015{\natexlab{b}})\citenamefont {Brito}, \citenamefont {Cardoso},
  and \citenamefont {Pani}}]{Brito:2014wla}%
  \BibitemOpen
  \bibfield  {author} {\bibinfo {author} {\bibfnamefont {R.}~\bibnamefont
  {Brito}}, \bibinfo {author} {\bibfnamefont {V.}~\bibnamefont {Cardoso}},  and
  \bibinfo {author} {\bibfnamefont {P.}~\bibnamefont {Pani}}, }\href {\doibase
  10.1088/0264-9381/32/13/134001} {\bibfield  {journal} {\bibinfo  {journal}
  {\emph {Class. Quant. Grav.}} }\textbf {\bibinfo {volume} {32}}, \bibinfo
  {pages} {134001} (\bibinfo {year} {2015}{\natexlab{b}})}, \Eprint
  {http://arxiv.org/abs/1411.0686} {arXiv:1411.0686}\BibitemShut {NoStop}%
\bibitem [{\citenamefont {Brito} \emph
  {et~al.}(2017{\natexlab{a}})\citenamefont {Brito}, \citenamefont {Ghosh},
  \citenamefont {Barausse}, \citenamefont {Berti}, \citenamefont {Cardoso},
  \citenamefont {Dvorkin}, \citenamefont {Klein}, and \citenamefont
  {Pani}}]{Brito:2017wnc}%
  \BibitemOpen
  \bibfield  {author} {\bibinfo {author} {\bibfnamefont {R.}~\bibnamefont
  {Brito}}, \bibinfo {author} {\bibfnamefont {S.}~\bibnamefont {Ghosh}},
  \bibinfo {author} {\bibfnamefont {E.}~\bibnamefont {Barausse}}, \bibinfo
  {author} {\bibfnamefont {E.}~\bibnamefont {Berti}}, \bibinfo {author}
  {\bibfnamefont {V.}~\bibnamefont {Cardoso}}, \bibinfo {author} {\bibfnamefont
  {I.}~\bibnamefont {Dvorkin}}, \bibinfo {author} {\bibfnamefont
  {A.}~\bibnamefont {Klein}},  and \bibinfo {author} {\bibfnamefont
  {P.}~\bibnamefont {Pani}}, }\href {\doibase 10.1103/PhysRevLett.119.131101}
  {\bibfield  {journal} {\bibinfo  {journal} {\emph {Phys. Rev. Lett.}}
  }\textbf {\bibinfo {volume} {119}}, \bibinfo {pages} {131101} (\bibinfo
  {year} {2017}{\natexlab{a}})}, \Eprint {http://arxiv.org/abs/1706.05097}
  {arXiv:1706.05097}\BibitemShut {NoStop}%
\bibitem [{\citenamefont {Brito} \emph
  {et~al.}(2017{\natexlab{b}})\citenamefont {Brito}, \citenamefont {Ghosh},
  \citenamefont {Barausse}, \citenamefont {Berti}, \citenamefont {Cardoso},
  \citenamefont {Dvorkin}, \citenamefont {Klein}, and \citenamefont
  {Pani}}]{Brito:2017zvb}%
  \BibitemOpen
  \bibfield  {author} {\bibinfo {author} {\bibfnamefont {R.}~\bibnamefont
  {Brito}}, \bibinfo {author} {\bibfnamefont {S.}~\bibnamefont {Ghosh}},
  \bibinfo {author} {\bibfnamefont {E.}~\bibnamefont {Barausse}}, \bibinfo
  {author} {\bibfnamefont {E.}~\bibnamefont {Berti}}, \bibinfo {author}
  {\bibfnamefont {V.}~\bibnamefont {Cardoso}}, \bibinfo {author} {\bibfnamefont
  {I.}~\bibnamefont {Dvorkin}}, \bibinfo {author} {\bibfnamefont
  {A.}~\bibnamefont {Klein}},  and \bibinfo {author} {\bibfnamefont
  {P.}~\bibnamefont {Pani}}, }\href {\doibase 10.1103/PhysRevD.96.064050}
  {\bibfield  {journal} {\bibinfo  {journal} {\emph {Phys. Rev.}} }\textbf
  {\bibinfo {volume} {D96}}, \bibinfo {pages} {064050} (\bibinfo {year}
  {2017}{\natexlab{b}})}, \Eprint {http://arxiv.org/abs/1706.06311}
  {arXiv:1706.06311}\BibitemShut {NoStop}%
\bibitem [{\citenamefont {Abbott} \emph {et~al.}(2016)}]{Abbott:2016blz}%
  \BibitemOpen
  \bibfield  {author} {\bibinfo {author} {\bibfnamefont {B.P.} \bibnamefont
  {Abbott}} \emph {et~al.} (\bibinfo {collaboration} {Virgo, LIGO Scientific}),
  }\href {\doibase 10.1103/PhysRevLett.116.061102} {\bibfield  {journal}
  {\bibinfo  {journal} {\emph {Phys. Rev. Lett.}} }\textbf {\bibinfo {volume}
  {116}}, \bibinfo {pages} {061102} (\bibinfo {year} {2016})}, \Eprint
  {http://arxiv.org/abs/1602.03837} {arXiv:1602.03837}\BibitemShut {NoStop}%
\bibitem [{\citenamefont {Cardoso} \emph
  {et~al.}(2016{\natexlab{b}})\citenamefont {Cardoso}, \citenamefont {Hopper},
  \citenamefont {Macedo}, \citenamefont {Palenzuela}, and \citenamefont
  {Pani}}]{Cardoso:2016oxy}%
  \BibitemOpen
  \bibfield  {author} {\bibinfo {author} {\bibfnamefont {V.}~\bibnamefont
  {Cardoso}}, \bibinfo {author} {\bibfnamefont {S.}~\bibnamefont {Hopper}},
  \bibinfo {author} {\bibfnamefont {C.F.B.} \bibnamefont {Macedo}}, \bibinfo
  {author} {\bibfnamefont {C.}~\bibnamefont {Palenzuela}},  and \bibinfo
  {author} {\bibfnamefont {P.}~\bibnamefont {Pani}}, }\href {\doibase
  10.1103/PhysRevD.94.084031} {\bibfield  {journal} {\bibinfo  {journal} {\emph
  {Phys. Rev.}} }\textbf {\bibinfo {volume} {D94}}, \bibinfo {pages} {084031}
  (\bibinfo {year} {2016}{\natexlab{b}})}, \Eprint
  {http://arxiv.org/abs/1608.08637} {arXiv:1608.08637}\BibitemShut {NoStop}%
\bibitem [{\citenamefont {Cardoso} \emph
  {et~al.}(2016{\natexlab{c}})\citenamefont {Cardoso}, \citenamefont {Franzin},
  and \citenamefont {Pani}}]{Cardoso:2016rao}%
  \BibitemOpen
  \bibfield  {author} {\bibinfo {author} {\bibfnamefont {V.}~\bibnamefont
  {Cardoso}}, \bibinfo {author} {\bibfnamefont {E.}~\bibnamefont {Franzin}},
  and \bibinfo {author} {\bibfnamefont {P.}~\bibnamefont {Pani}}, }\href
  {\doibase 10.1103/PhysRevLett.117.089902, 10.1103/PhysRevLett.116.171101}
  {\bibfield  {journal} {\bibinfo  {journal} {\emph {Phys. Rev. Lett.}}
  }\textbf {\bibinfo {volume} {116}}, \bibinfo {pages} {171101} (\bibinfo
  {year} {2016}{\natexlab{c}})}, \bibinfo {note} {[Erratum: Phys. Rev.
  Lett.117,no.8,089902(2016)]}, \Eprint {http://arxiv.org/abs/1602.07309}
  {arXiv:1602.07309}\BibitemShut {NoStop}%
\bibitem [{\citenamefont {Rosa} and \citenamefont
  {Kephart}(2017)}]{Rosa:2017ury}%
  \BibitemOpen
  \bibfield  {author} {\bibinfo {author} {\bibfnamefont {J.G.} \bibnamefont
  {Rosa}} and \bibinfo {author} {\bibfnamefont {T.W.} \bibnamefont {Kephart}},
  }\href@noop {} {  (\bibinfo {year} {2017})}, \Eprint
  {http://arxiv.org/abs/1709.06581} {arXiv:1709.06581}\BibitemShut {NoStop}%
\bibitem [{\citenamefont {Lee}(1977)}]{Lee:1977gk}%
  \BibitemOpen
  \bibfield  {author} {\bibinfo {author} {\bibfnamefont {C.H.} \bibnamefont
  {Lee}}, }\href {\doibase 10.1016/0370-2693(77)90189-7} {\bibfield  {journal}
  {\bibinfo  {journal} {\emph {Phys. Lett.}} }\textbf {\bibinfo {volume}
  {B68}}, \bibinfo {pages} {152} (\bibinfo {year} {1977})}\BibitemShut
  {NoStop}%
\bibitem [{\citenamefont {Unruh}(1973)}]{Unruh:1973bda}%
  \BibitemOpen
  \bibfield  {author} {\bibinfo {author} {\bibfnamefont {W.}~\bibnamefont
  {Unruh}}, }\href {\doibase 10.1103/PhysRevLett.31.1265} {\bibfield  {journal}
  {\bibinfo  {journal} {\emph {Phys. Rev. Lett.}} }\textbf {\bibinfo {volume}
  {31}}, \bibinfo {pages} {1265} (\bibinfo {year} {1973})}\BibitemShut
  {NoStop}%
\bibitem [{\citenamefont {Chandrasekhar}(1976)}]{Chandrasekhar:1976ap}%
  \BibitemOpen
  \bibfield  {author} {\bibinfo {author} {\bibfnamefont {S.}~\bibnamefont
  {Chandrasekhar}}, }\href {\doibase 10.1098/rspa.1976.0090} {\bibfield
  {journal} {\bibinfo  {journal} {\emph {Proc. Roy. Soc. Lond.}} }\textbf
  {\bibinfo {volume} {A349}}, \bibinfo {pages} {571} (\bibinfo {year}
  {1976})}\BibitemShut {NoStop}%
\bibitem [{\citenamefont {Page}(1976)}]{Page:1976jj}%
  \BibitemOpen
  \bibfield  {author} {\bibinfo {author} {\bibfnamefont {D.N.} \bibnamefont
  {Page}}, }\href {\doibase 10.1103/PhysRevD.14.1509} {\bibfield  {journal}
  {\bibinfo  {journal} {\emph {Phys. Rev.}} }\textbf {\bibinfo {volume} {D14}},
  \bibinfo {pages} {1509} (\bibinfo {year} {1976})}\BibitemShut {NoStop}%
\bibitem [{\citenamefont {Bekenstein}(1973)}]{Bekenstein:1973mi}%
  \BibitemOpen
  \bibfield  {author} {\bibinfo {author} {\bibfnamefont {J.D.} \bibnamefont
  {Bekenstein}}, }\href {\doibase 10.1103/PhysRevD.7.949} {\bibfield  {journal}
  {\bibinfo  {journal} {\emph {Phys. Rev.}} }\textbf {\bibinfo {volume} {D7}},
  \bibinfo {pages} {949} (\bibinfo {year} {1973})}\BibitemShut {NoStop}%
\bibitem [{\citenamefont {Hawking} and \citenamefont
  {Ellis}(2011)}]{Hawking:1973uf}%
  \BibitemOpen
  \bibfield  {author} {\bibinfo {author} {\bibfnamefont {S.W.} \bibnamefont
  {Hawking}} and \bibinfo {author} {\bibfnamefont {G.F.R.} \bibnamefont
  {Ellis}}, }\href {\doibase 10.1017/CBO9780511524646} {\emph {\bibinfo {title}
  {{The Large Scale Structure of Space-Time}}}}, Cambridge Monographs on
  Mathematical Physics (\bibinfo  {publisher} {Cambridge University Press},
  \bibinfo {year} {2011})\BibitemShut {NoStop}%
\bibitem [{\citenamefont {Natario} \emph {et~al.}(2016)\citenamefont {Natario},
  \citenamefont {Queimada}, and \citenamefont {Vicente}}]{Natario:2016bay}%
  \BibitemOpen
  \bibfield  {author} {\bibinfo {author} {\bibfnamefont {J.}~\bibnamefont
  {Natario}}, \bibinfo {author} {\bibfnamefont {L.}~\bibnamefont {Queimada}},
  and \bibinfo {author} {\bibfnamefont {R.}~\bibnamefont {Vicente}}, }\href
  {\doibase 10.1088/0264-9381/33/17/175002} {\bibfield  {journal} {\bibinfo
  {journal} {\emph {Class. Quant. Grav.}} }\textbf {\bibinfo {volume} {33}},
  \bibinfo {pages} {175002} (\bibinfo {year} {2016})}, \Eprint
  {http://arxiv.org/abs/1601.06809} {arXiv:1601.06809}\BibitemShut {NoStop}%
\bibitem [{\citenamefont {T\'oth}(2016)}]{Toth:2015cda}%
  \BibitemOpen
  \bibfield  {author} {\bibinfo {author} {\bibfnamefont {G.Z.} \bibnamefont
  {T\'oth}}, }\href {\doibase 10.1088/0264-9381/33/11/115012} {\bibfield
  {journal} {\bibinfo  {journal} {\emph {Class. Quant. Grav.}} }\textbf
  {\bibinfo {volume} {33}}, \bibinfo {pages} {115012} (\bibinfo {year}
  {2016})}, \Eprint {http://arxiv.org/abs/1509.02878}
  {arXiv:1509.02878}\BibitemShut {NoStop}%
\bibitem [{\citenamefont {Misner} \emph {et~al.}(1973)\citenamefont {Misner},
  \citenamefont {Thorne}, and \citenamefont {Wheeler}}]{gravitation}%
  \BibitemOpen
  \bibfield  {author} {\bibinfo {author} {\bibfnamefont {C.W.} \bibnamefont
  {Misner}}, \bibinfo {author} {\bibfnamefont {K.P.} \bibnamefont {Thorne}},
  and \bibinfo {author} {\bibfnamefont {J.A.} \bibnamefont {Wheeler}}, }\href
  {\doibase 10.1017/CBO9780511524646} {\emph {\bibinfo {title}
  {{Gravitation}}}} (\bibinfo  {publisher} {W. H. Freeman San Francisco},
  \bibinfo {year} {1973})\BibitemShut {NoStop}%
\bibitem [{\citenamefont {Richartz} \emph {et~al.}(2009)\citenamefont
  {Richartz}, \citenamefont {Weinfurtner}, \citenamefont {Penner}, and
  \citenamefont {Unruh}}]{Richartz:2009mi}%
  \BibitemOpen
  \bibfield  {author} {\bibinfo {author} {\bibfnamefont {M.}~\bibnamefont
  {Richartz}}, \bibinfo {author} {\bibfnamefont {S.}~\bibnamefont
  {Weinfurtner}}, \bibinfo {author} {\bibfnamefont {A.J.} \bibnamefont
  {Penner}},  and \bibinfo {author} {\bibfnamefont {W.G.} \bibnamefont
  {Unruh}}, }\href {\doibase 10.1103/PhysRevD.80.124016} {\bibfield  {journal}
  {\bibinfo  {journal} {\emph {Phys. Rev.}} }\textbf {\bibinfo {volume} {D80}},
  \bibinfo {pages} {124016} (\bibinfo {year} {2009})}, \Eprint
  {http://arxiv.org/abs/0909.2317} {arXiv:0909.2317}\BibitemShut {NoStop}%
\bibitem [{\citenamefont {Penrose} and \citenamefont
  {Floyd}(1971)}]{Penrose:1971uk}%
  \BibitemOpen
  \bibfield  {author} {\bibinfo {author} {\bibfnamefont {R.}~\bibnamefont
  {Penrose}} and \bibinfo {author} {\bibfnamefont {R.M.} \bibnamefont {Floyd}},
  }\href@noop {} {\bibfield  {journal} {\bibinfo  {journal} {\emph {Nature}}
  }\textbf {\bibinfo {volume} {229}}, \bibinfo {pages} {177} (\bibinfo {year}
  {1971})}\BibitemShut {NoStop}%
\bibitem [{\citenamefont {Denardo} and \citenamefont
  {Ruffini}(1973)}]{Denardo:1973pyo}%
  \BibitemOpen
  \bibfield  {author} {\bibinfo {author} {\bibfnamefont {G.}~\bibnamefont
  {Denardo}} and \bibinfo {author} {\bibfnamefont {R.}~\bibnamefont {Ruffini}},
  }\href {\doibase 10.1016/0370-2693(73)90198-6} {\bibfield  {journal}
  {\bibinfo  {journal} {\emph {Phys. Lett.}} }\textbf {\bibinfo {volume}
  {B45}}, \bibinfo {pages} {259} (\bibinfo {year} {1973})}\BibitemShut
  {NoStop}%
\bibitem [{\citenamefont {{Bhat}} \emph {et~al.}(1985)\citenamefont {{Bhat}},
  \citenamefont {{Dhurandhar}}, and \citenamefont
  {{Dadhich}}}]{1985JApA....6...85B}%
  \BibitemOpen
  \bibfield  {author} {\bibinfo {author} {\bibfnamefont {M.}~\bibnamefont
  {{Bhat}}}, \bibinfo {author} {\bibfnamefont {S.}~\bibnamefont
  {{Dhurandhar}}},  and \bibinfo {author} {\bibfnamefont {N.}~\bibnamefont
  {{Dadhich}}}, }\href {\doibase 10.1007/BF02715080} {\bibfield  {journal}
  {\bibinfo  {journal} {\emph {Journal of Astrophysics and Astronomy}} }\textbf
  {\bibinfo {volume} {6}}, \bibinfo {pages} {85} (\bibinfo {year}
  {1985})}\BibitemShut {NoStop}%
\bibitem [{\citenamefont {Cardoso} and \citenamefont
  {Pani}(2017{\natexlab{a}})}]{Cardoso:2017cqb}%
  \BibitemOpen
  \bibfield  {author} {\bibinfo {author} {\bibfnamefont {V.}~\bibnamefont
  {Cardoso}} and \bibinfo {author} {\bibfnamefont {P.}~\bibnamefont {Pani}},
  }\href {\doibase 10.1038/s41550-017-0225-y} {\bibfield  {journal} {\bibinfo
  {journal} {\emph {Nat. Astron.}} }\textbf {\bibinfo {volume} {1}}, \bibinfo
  {pages} {586} (\bibinfo {year} {2017}{\natexlab{a}})}, \Eprint
  {http://arxiv.org/abs/1709.01525} {arXiv:1709.01525}\BibitemShut {NoStop}%
\bibitem [{\citenamefont {Cardoso} and \citenamefont
  {Pani}(2017{\natexlab{b}})}]{Cardoso:2017njb}%
  \BibitemOpen
  \bibfield  {author} {\bibinfo {author} {\bibfnamefont {V.}~\bibnamefont
  {Cardoso}} and \bibinfo {author} {\bibfnamefont {P.}~\bibnamefont {Pani}},
  }\href@noop {} {  (\bibinfo {year} {2017}{\natexlab{b}})}, \Eprint
  {http://arxiv.org/abs/1707.03021} {arXiv:1707.03021}\BibitemShut {NoStop}%
\bibitem [{\citenamefont {Mark} \emph {et~al.}(2017)\citenamefont {Mark},
  \citenamefont {Zimmerman}, \citenamefont {Du}, and \citenamefont
  {Chen}}]{Mark:2017dnq}%
  \BibitemOpen
  \bibfield  {author} {\bibinfo {author} {\bibfnamefont {Z.}~\bibnamefont
  {Mark}}, \bibinfo {author} {\bibfnamefont {A.}~\bibnamefont {Zimmerman}},
  \bibinfo {author} {\bibfnamefont {S.M.} \bibnamefont {Du}},  and \bibinfo
  {author} {\bibfnamefont {Y.}~\bibnamefont {Chen}}, }\href {\doibase
  10.1103/PhysRevD.96.084002} {\bibfield  {journal} {\bibinfo  {journal} {\emph
  {Phys. Rev.}} }\textbf {\bibinfo {volume} {D96}}, \bibinfo {pages} {084002}
  (\bibinfo {year} {2017})}, \Eprint {http://arxiv.org/abs/1706.06155}
  {arXiv:1706.06155}\BibitemShut {NoStop}%
\bibitem [{\citenamefont {Hansen} and \citenamefont
  {Ravndal}(1981)}]{Hansen:1980nc}%
  \BibitemOpen
  \bibfield  {author} {\bibinfo {author} {\bibfnamefont {A.}~\bibnamefont
  {Hansen}} and \bibinfo {author} {\bibfnamefont {F.}~\bibnamefont {Ravndal}},
  }\href {\doibase 10.1088/0031-8949/23/6/002} {\bibfield  {journal} {\bibinfo
  {journal} {\emph {Phys. Scripta}} }\textbf {\bibinfo {volume} {23}}, \bibinfo
  {pages} {1036} (\bibinfo {year} {1981})}\BibitemShut {NoStop}%
\bibitem [{\citenamefont {Nambu} and \citenamefont
  {Jona-Lasinio}(1961)}]{Nambu}%
  \BibitemOpen
  \bibfield  {author} {\bibinfo {author} {\bibfnamefont {Y.}~\bibnamefont
  {Nambu}} and \bibinfo {author} {\bibfnamefont {G.}~\bibnamefont
  {Jona-Lasinio}}, }\href {\doibase 10.1103/PhysRev.122.345} {\bibfield
  {journal} {\bibinfo  {journal} {\emph {Phys. Rev.}} }\textbf {\bibinfo
  {volume} {122}}, \bibinfo {pages} {345} (\bibinfo {year} {1961})}\BibitemShut
  {NoStop}%
\bibitem [{\citenamefont {Detweiler}(1980)}]{Detweiler:1980uk}%
  \BibitemOpen
  \bibfield  {author} {\bibinfo {author} {\bibfnamefont {S.L.} \bibnamefont
  {Detweiler}}, }\href {\doibase 10.1103/PhysRevD.22.2323} {\bibfield
  {journal} {\bibinfo  {journal} {\emph {Phys. Rev.}} }\textbf {\bibinfo
  {volume} {D22}}, \bibinfo {pages} {2323} (\bibinfo {year}
  {1980})}\BibitemShut {NoStop}%
\bibitem [{\citenamefont {Cardoso} and \citenamefont
  {Dias}(2004)}]{Cardoso:2004hs}%
  \BibitemOpen
  \bibfield  {author} {\bibinfo {author} {\bibfnamefont {V.}~\bibnamefont
  {Cardoso}} and \bibinfo {author} {\bibfnamefont {O.J.C.} \bibnamefont
  {Dias}}, }\href {\doibase 10.1103/PhysRevD.70.084011} {\bibfield  {journal}
  {\bibinfo  {journal} {\emph {Phys. Rev.}} }\textbf {\bibinfo {volume} {D70}},
  \bibinfo {pages} {084011} (\bibinfo {year} {2004})}, \Eprint
  {http://arxiv.org/abs/hep-th/0405006} {arXiv:hep-th/0405006}\BibitemShut
  {NoStop}%
\bibitem [{\citenamefont {Press} and \citenamefont
  {Teukolsky}(1972)}]{Press:1972zz}%
  \BibitemOpen
  \bibfield  {author} {\bibinfo {author} {\bibfnamefont {W.H.} \bibnamefont
  {Press}} and \bibinfo {author} {\bibfnamefont {S.A.} \bibnamefont
  {Teukolsky}}, }\href {\doibase 10.1038/238211a0} {\bibfield  {journal}
  {\bibinfo  {journal} {\emph {Nature}} }\textbf {\bibinfo {volume} {238}},
  \bibinfo {pages} {211} (\bibinfo {year} {1972})}\BibitemShut {NoStop}%
\bibitem [{\citenamefont {Cardoso} \emph {et~al.}(2004)\citenamefont {Cardoso},
  \citenamefont {Dias}, \citenamefont {Lemos}, and \citenamefont
  {Yoshida}}]{Cardoso:2004nk}%
  \BibitemOpen
  \bibfield  {author} {\bibinfo {author} {\bibfnamefont {V.}~\bibnamefont
  {Cardoso}}, \bibinfo {author} {\bibfnamefont {O.J.C.} \bibnamefont {Dias}},
  \bibinfo {author} {\bibfnamefont {J.P.S.} \bibnamefont {Lemos}},  and
  \bibinfo {author} {\bibfnamefont {S.}~\bibnamefont {Yoshida}}, }\href
  {\doibase 10.1103/PhysRevD.70.049903, 10.1103/PhysRevD.70.044039} {\bibfield
  {journal} {\bibinfo  {journal} {\emph {Phys. Rev.}} }\textbf {\bibinfo
  {volume} {D70}}, \bibinfo {pages} {044039} (\bibinfo {year} {2004})},
  \bibinfo {note} {[Erratum: Phys. Rev.D70,049903(2004)]}, \Eprint
  {http://arxiv.org/abs/hep-th/0404096} {arXiv:hep-th/0404096}\BibitemShut
  {NoStop}%
\bibitem [{\citenamefont {Vilenkin}(1978)}]{VILENKIN1978301}%
  \BibitemOpen
  \bibfield  {author} {\bibinfo {author} {\bibfnamefont {A.}~\bibnamefont
  {Vilenkin}}, }\href {\doibase https://doi.org/10.1016/0370-2693(78)90027-8}
  {\bibfield  {journal} {\bibinfo  {journal} {\emph {Physics Letters B}}
  }\textbf {\bibinfo {volume} {78}}, \bibinfo {pages} {301 } (\bibinfo {year}
  {1978})}\BibitemShut {NoStop}%
\bibitem [{\citenamefont {Chesler} and \citenamefont
  {Lowe}(2018)}]{Chesler:2018txn}%
  \BibitemOpen
  \bibfield  {author} {\bibinfo {author} {\bibfnamefont {P.M.} \bibnamefont
  {Chesler}} and \bibinfo {author} {\bibfnamefont {D.A.} \bibnamefont {Lowe}},
  }\href@noop {} {  (\bibinfo {year} {2018})}, \Eprint
  {http://arxiv.org/abs/1801.09711} {arXiv:1801.09711}\BibitemShut {NoStop}%
\bibitem [{\citenamefont {Arvanitaki} and \citenamefont
  {Dubovsky}(2011)}]{Arvanitaki:2010sy}%
  \BibitemOpen
  \bibfield  {author} {\bibinfo {author} {\bibfnamefont {A.}~\bibnamefont
  {Arvanitaki}} and \bibinfo {author} {\bibfnamefont {S.}~\bibnamefont
  {Dubovsky}}, }\href {\doibase 10.1103/PhysRevD.83.044026} {\bibfield
  {journal} {\bibinfo  {journal} {\emph {Phys. Rev.}} }\textbf {\bibinfo
  {volume} {D83}}, \bibinfo {pages} {044026} (\bibinfo {year} {2011})}, \Eprint
  {http://arxiv.org/abs/1004.3558} {arXiv:1004.3558}\BibitemShut {NoStop}%
\bibitem [{\citenamefont {Pani} \emph {et~al.}(2012)\citenamefont {Pani},
  \citenamefont {Cardoso}, \citenamefont {Gualtieri}, \citenamefont {Berti},
  and \citenamefont {Ishibashi}}]{Pani:2012vp}%
  \BibitemOpen
  \bibfield  {author} {\bibinfo {author} {\bibfnamefont {P.}~\bibnamefont
  {Pani}}, \bibinfo {author} {\bibfnamefont {V.}~\bibnamefont {Cardoso}},
  \bibinfo {author} {\bibfnamefont {L.}~\bibnamefont {Gualtieri}}, \bibinfo
  {author} {\bibfnamefont {E.}~\bibnamefont {Berti}},  and \bibinfo {author}
  {\bibfnamefont {A.}~\bibnamefont {Ishibashi}}, }\href {\doibase
  10.1103/PhysRevLett.109.131102} {\bibfield  {journal} {\bibinfo  {journal}
  {\emph {Phys. Rev. Lett.}} }\textbf {\bibinfo {volume} {109}}, \bibinfo
  {pages} {131102} (\bibinfo {year} {2012})}, \Eprint
  {http://arxiv.org/abs/1209.0465} {arXiv:1209.0465}\BibitemShut {NoStop}%
\bibitem [{\citenamefont {Brito} \emph {et~al.}(2013)\citenamefont {Brito},
  \citenamefont {Cardoso}, and \citenamefont {Pani}}]{Brito:2013wya}%
  \BibitemOpen
  \bibfield  {author} {\bibinfo {author} {\bibfnamefont {R.}~\bibnamefont
  {Brito}}, \bibinfo {author} {\bibfnamefont {V.}~\bibnamefont {Cardoso}},  and
  \bibinfo {author} {\bibfnamefont {P.}~\bibnamefont {Pani}}, }\href {\doibase
  10.1103/PhysRevD.88.023514} {\bibfield  {journal} {\bibinfo  {journal} {\emph
  {Phys. Rev.}} }\textbf {\bibinfo {volume} {D88}}, \bibinfo {pages} {023514}
  (\bibinfo {year} {2013})}, \Eprint {http://arxiv.org/abs/1304.6725}
  {arXiv:1304.6725}\BibitemShut {NoStop}%
\bibitem [{\citenamefont {Hod}(2012)}]{Hod:2012px}%
  \BibitemOpen
  \bibfield  {author} {\bibinfo {author} {\bibfnamefont {S.}~\bibnamefont
  {Hod}}, }\href {\doibase 10.1103/PhysRevD.86.129902,
  10.1103/PhysRevD.86.104026} {\bibfield  {journal} {\bibinfo  {journal} {\emph
  {Phys. Rev.}} }\textbf {\bibinfo {volume} {D86}}, \bibinfo {pages} {104026}
  (\bibinfo {year} {2012})}, \bibinfo {note} {[Erratum: Phys.
  Rev.D86,129902(2012)]}, \Eprint {http://arxiv.org/abs/1211.3202}
  {arXiv:1211.3202}\BibitemShut {NoStop}%
\bibitem [{\citenamefont {Herdeiro} and \citenamefont
  {Radu}(2014)}]{Herdeiro:2014goa}%
  \BibitemOpen
  \bibfield  {author} {\bibinfo {author} {\bibfnamefont {C.A.R.} \bibnamefont
  {Herdeiro}} and \bibinfo {author} {\bibfnamefont {E.}~\bibnamefont {Radu}},
  }\href {\doibase 10.1103/PhysRevLett.112.221101} {\bibfield  {journal}
  {\bibinfo  {journal} {\emph {Phys. Rev. Lett.}} }\textbf {\bibinfo {volume}
  {112}}, \bibinfo {pages} {221101} (\bibinfo {year} {2014})}, \Eprint
  {http://arxiv.org/abs/1403.2757} {arXiv:1403.2757}\BibitemShut {NoStop}%
\bibitem [{\citenamefont {Herdeiro} and \citenamefont
  {Radu}(2015)}]{Herdeiro:2015waa}%
  \BibitemOpen
  \bibfield  {author} {\bibinfo {author} {\bibfnamefont {C.A.R.} \bibnamefont
  {Herdeiro}} and \bibinfo {author} {\bibfnamefont {E.}~\bibnamefont {Radu}},
  }\bibfield  {booktitle} {\emph {\bibinfo {booktitle} {{Proceedings, 7th Black
  Holes Workshop 2014: Aveiro, Portugal, December 18-19, 2014}}}, }\href
  {\doibase 10.1142/S0218271815420146} {\bibfield  {journal} {\bibinfo
  {journal} {\emph {Int. J. Mod. Phys.}} }\textbf {\bibinfo {volume} {D24}},
  \bibinfo {pages} {1542014} (\bibinfo {year} {2015})}, \Eprint
  {http://arxiv.org/abs/1504.08209} {arXiv:1504.08209}\BibitemShut {NoStop}%
\bibitem [{\citenamefont {Newman} and \citenamefont
  {Penrose}(1962)}]{Newman:1961qr}%
  \BibitemOpen
  \bibfield  {author} {\bibinfo {author} {\bibfnamefont {E.}~\bibnamefont
  {Newman}} and \bibinfo {author} {\bibfnamefont {R.}~\bibnamefont {Penrose}},
  }\href {\doibase 10.1063/1.1724257} {\bibfield  {journal} {\bibinfo
  {journal} {\emph {J. Math. Phys.}} }\textbf {\bibinfo {volume} {3}}, \bibinfo
  {pages} {566} (\bibinfo {year} {1962})}\BibitemShut {NoStop}%
\bibitem [{\citenamefont {Banados} and \citenamefont
  {Reyes}(2016)}]{Banados:2016zim}%
  \BibitemOpen
  \bibfield  {author} {\bibinfo {author} {\bibfnamefont {M.}~\bibnamefont
  {Banados}} and \bibinfo {author} {\bibfnamefont {I.A.} \bibnamefont {Reyes}},
  }\href {\doibase 10.1142/S0218271816300214} {\bibfield  {journal} {\bibinfo
  {journal} {\emph {Int. J. Mod. Phys.}} }\textbf {\bibinfo {volume} {D25}},
  \bibinfo {pages} {1630021} (\bibinfo {year} {2016})}, \Eprint
  {http://arxiv.org/abs/1601.03616} {arXiv:1601.03616}\BibitemShut {NoStop}%
\bibitem [{\citenamefont {Klein}(1929)}]{Klein}%
  \BibitemOpen
  \bibfield  {author} {\bibinfo {author} {\bibfnamefont {O.}~\bibnamefont
  {Klein}}, }\href@noop {} {\bibfield  {journal} {\bibinfo  {journal} {\emph
  {Zeitschrift fr Physik}} }\textbf {\bibinfo {volume} {\textbf{53}}}, \bibinfo
  {pages} {157} (\bibinfo {year} {1929})}\BibitemShut {NoStop}%
\bibitem [{\citenamefont {Sauter}(1931)}]{Sauter:1931zz}%
  \BibitemOpen
  \bibfield  {author} {\bibinfo {author} {\bibfnamefont {F.}~\bibnamefont
  {Sauter}}, }\href {\doibase 10.1007/BF01339461} {\bibfield  {journal}
  {\bibinfo  {journal} {\emph {Z. Phys.}} }\textbf {\bibinfo {volume} {69}},
  \bibinfo {pages} {742} (\bibinfo {year} {1931})}\BibitemShut {NoStop}%
\bibitem [{\citenamefont {Calogeracos} and \citenamefont
  {Dombey}(1999)}]{Calogeracos:1999yp}%
  \BibitemOpen
  \bibfield  {author} {\bibinfo {author} {\bibfnamefont {A.}~\bibnamefont
  {Calogeracos}} and \bibinfo {author} {\bibfnamefont {N.}~\bibnamefont
  {Dombey}}, }\href {\doibase 10.1080/001075199181387} {\bibfield  {journal}
  {\bibinfo  {journal} {\emph {Contemp. Phys.}} }\textbf {\bibinfo {volume}
  {40}}, \bibinfo {pages} {313} (\bibinfo {year} {1999})}, \Eprint
  {http://arxiv.org/abs/quant-ph/9905076} {arXiv:quant-ph/9905076}\BibitemShut
  {NoStop}%
\bibitem [{\citenamefont {Damour}(1975)}]{Damour:1975pr}%
  \BibitemOpen
  \bibfield  {author} {\bibinfo {author} {\bibfnamefont {T.}~\bibnamefont
  {Damour}}, }in \href@noop {} {\emph {\bibinfo {booktitle} {{Marcel Grossmann
  Meeting on the Recent Progress of the Fundamentals of General Relativity
  Trieste, Italy, July 7-12, 1975}}}} (\bibinfo {year} {1975}) pp. \bibinfo
  {pages} {459--482}\BibitemShut {NoStop}%
\bibitem [{\citenamefont {Arderucio}(2014)}]{Arderucio:2014oua}%
  \BibitemOpen
  \bibfield  {author} {\bibinfo {author} {\bibfnamefont {B.}~\bibnamefont
  {Arderucio}}, }\href@noop {} {  (\bibinfo {year} {2014})}, \Eprint
  {http://arxiv.org/abs/1404.3421} {arXiv:1404.3421}\BibitemShut {NoStop}%
\bibitem [{\citenamefont {Soler}(1970)}]{SOLER}%
  \BibitemOpen
  \bibfield  {author} {\bibinfo {author} {\bibfnamefont {M.}~\bibnamefont
  {Soler}}, }\href {\doibase 10.1103/PhysRevD.1.2766} {\bibfield  {journal}
  {\bibinfo  {journal} {\emph {Phys. Rev. D}} }\textbf {\bibinfo {volume} {1}},
  \bibinfo {pages} {2766} (\bibinfo {year} {1970})}\BibitemShut {NoStop}%
\bibitem [{\citenamefont {Mosta} \emph {et~al.}(2010)\citenamefont {Mosta},
  \citenamefont {Palenzuela}, \citenamefont {Rezzolla}, \citenamefont {Lehner},
  \citenamefont {Yoshida}, and \citenamefont {Pollney}}]{Mosta:2009rr}%
  \BibitemOpen
  \bibfield  {author} {\bibinfo {author} {\bibfnamefont {P.}~\bibnamefont
  {Mosta}}, \bibinfo {author} {\bibfnamefont {C.}~\bibnamefont {Palenzuela}},
  \bibinfo {author} {\bibfnamefont {L.}~\bibnamefont {Rezzolla}}, \bibinfo
  {author} {\bibfnamefont {L.}~\bibnamefont {Lehner}}, \bibinfo {author}
  {\bibfnamefont {S.}~\bibnamefont {Yoshida}},  and \bibinfo {author}
  {\bibfnamefont {D.}~\bibnamefont {Pollney}}, }\href {\doibase
  10.1103/PhysRevD.81.064017} {\bibfield  {journal} {\bibinfo  {journal} {\emph
  {Phys. Rev.}} }\textbf {\bibinfo {volume} {D81}}, \bibinfo {pages} {064017}
  (\bibinfo {year} {2010})}, \Eprint {http://arxiv.org/abs/0912.2330}
  {arXiv:0912.2330}\BibitemShut {NoStop}%
\bibitem [{\citenamefont {Finster} \emph {et~al.}(1999)\citenamefont {Finster},
  \citenamefont {Smoller}, and \citenamefont {Yau}}]{Yau2}%
  \BibitemOpen
  \bibfield  {author} {\bibinfo {author} {\bibfnamefont {F.}~\bibnamefont
  {Finster}}, \bibinfo {author} {\bibfnamefont {J.}~\bibnamefont {Smoller}},
  and \bibinfo {author} {\bibfnamefont {S.T.} \bibnamefont {Yau}}, }\href
  {\doibase 10.1103/PhysRevD.59.104020} {\bibfield  {journal} {\bibinfo
  {journal} {\emph {Phys. Rev. D}} }\textbf {\bibinfo {volume} {59}}, \bibinfo
  {pages} {104020} (\bibinfo {year} {1999})}\BibitemShut {NoStop}%
\bibitem [{\citenamefont {Biedenharn} and \citenamefont
  {Louck}(1984)}]{Biedenharn_Louck_1984}%
  \BibitemOpen
  \bibfield  {author} {\bibinfo {author} {\bibfnamefont {L.C.} \bibnamefont
  {Biedenharn}} and \bibinfo {author} {\bibfnamefont {J.D.} \bibnamefont
  {Louck}}, }\href {\doibase 10.1017/CBO9780511759888} {\emph {\bibinfo {title}
  {Angular Momentum in Quantum Physics: Theory and Application}}}, Encyclopedia
  of Mathematics and its Applications (\bibinfo  {publisher} {Cambridge
  University Press}, \bibinfo {year} {1984})\BibitemShut {NoStop}%
\bibitem [{\citenamefont {Finster} \emph {et~al.}(2000)\citenamefont {Finster},
  \citenamefont {Smoller}, and \citenamefont {Yau}}]{Finster:1998ak}%
  \BibitemOpen
  \bibfield  {author} {\bibinfo {author} {\bibfnamefont {F.}~\bibnamefont
  {Finster}}, \bibinfo {author} {\bibfnamefont {J.}~\bibnamefont {Smoller}},
  and \bibinfo {author} {\bibfnamefont {S.T.} \bibnamefont {Yau}}, }\href
  {\doibase 10.1063/1.533234} {\bibfield  {journal} {\bibinfo  {journal} {\emph
  {J. Math. Phys.}} }\textbf {\bibinfo {volume} {41}}, \bibinfo {pages} {2173}
  (\bibinfo {year} {2000})}, \Eprint {http://arxiv.org/abs/gr-qc/9805050}
  {arXiv:gr-qc/9805050}\BibitemShut {NoStop}%
\end{thebibliography}

\end{document}